 \documentclass[iop]{emulateapj}
 
\bibstyle{ApJ}
\shorttitle{RIOTS4 Survey}
\shortauthors{Lamb, J. B.,  et al.}

\usepackage{epsfig}

\begin{document}

\newcommand{\Mcl}{M_{\rm cl}}
\newcommand{\Mcllo}{M_{\rm cl,lo}}
\newcommand{\mlo}{m_{\rm lo}}
\newcommand{\mhi}{m_{\rm hi}}
\newcommand{\mup}{m_{\rm up}}
\newcommand{\Nlo}{N_{*,{\rm lo}}}
\newcommand{\mmax}{m_{\rm max}}
\newcommand{\mmaxtwo}{m_{\rm max,2}}
\newcommand{\mratio}{m_{\rm max,2} / m_{\rm max}}

\slugcomment{***Accepted for publication in ApJ  ***}


\title{The Runaways and Isolated O-Type Star Spectroscopic Survey of the SMC (RIOTS4)\footnotemark[*]}\footnotetext[*]{This paper includes data gathered with the 6.5 meter Magellan Telescopes located at Las Campanas Observatory, Chile.}


\author{J. B. Lamb\altaffilmark{1,2}, M. S. Oey\altaffilmark{1},
D. M. Segura-Cox\altaffilmark{1,3}, A. S. Graus\altaffilmark{1,4}, 
D. C. Kiminki\altaffilmark{5}, J. B. Golden-Marx\altaffilmark{1}, \\ and J. Wm. Parker\altaffilmark{6}}

\altaffiltext{1}{Astronomy Department, University of Michigan, 1085 S. University Ave.,
  Ann Arbor, MI 48109-1107}
\altaffiltext{2}{Department of Physical Sciences, Nassau Community College, One Education Drive, Garden City, NY 11530}
\altaffiltext{3}{Department of Astronomy, University of Illinois, Urbana, IL 61801}
\altaffiltext{4}{Department of Physics and Astronomy, University of California, Irvine, CA 92697}
\altaffiltext{5}{Department of Astronomy, University of Arizona, Tucson, AZ 85721}
\altaffiltext{6}{Southwest Research Institute, Department of Space Studies, Suite 300, 1050 Walnut Street, Boulder, CO 80302-5150, USA}



\begin{abstract}
We present the Runaways and Isolated O-Type Star Spectroscopic Survey of the SMC
(RIOTS4), a spatially complete survey of uniformly selected field OB stars
that covers the entire star-forming body of the SMC.  Using
the IMACS multislit spectrograph and MIKE echelle spectrograph on the
Magellan telescopes, we obtained spectra of 374 early-type field
stars that are at least 28 pc from any other OB candidates.  
We also obtained spectra of an additional 23 field stars in the SMC bar
identified from slightly different photometric criteria.
Here, we present the observational catalog of stars in the RIOTS4
survey, including spectral classifications and radial velocities.
For three multi-slit fields covering 8\% of our sample, we carried out
monitoring observations over 9--16 epochs to study binarity, finding
a spectroscopic, massive binary frequency of at least $\sim$60\% in this subsample.
Classical Oe/Be stars represent a large fraction of RIOTS4 (42\%), 
occurring at much higher frequency than in the Galaxy, consistent with
expectation at low metallicity.
RIOTS4 confirmed a steep upper IMF in the field, apparently
caused by the inability of the most massive stars to form in the
smallest clusters.  Our survey also yields evidence for in-situ field OB star
formation, and properties of field emission-line star populations,
including sgB[e] stars and classical Oe/Be stars.
We also discuss the radial velocity distribution and its
relation to SMC kinematics and runaway stars.  
RIOTS4 presents a first quantitative characterization of field OB
stars in an external galaxy, including the contributions of
sparse, but normal, star formation; runaway stars; and candidate
isolated star formation.  
\end{abstract}






\section{Introduction}
The standard model of star formation has been that most, if not all, stars form in clusters \citep[e.g.,][]{Lada03}, with the most
massive stars aggregating in the dense cores of clusters.  It is
intuitive that massive O stars form preferentially from the plentiful
gas reservoirs of giant molecular clouds (GMCs).  However, another
significant population of massive stars exists in an environment of
the opposite extreme.  These massive stars are far removed from dense
clusters or OB associations and instead appear isolated within the
sparse field population.  The physical properties and origin of this
field massive star population remain unclear, despite the fact that it accounts for
20--30\% of the massive stars in star-forming galaxies \citep{Oey04}.   The
existence of such stars in isolation poses a challenge for theories of massive star
formation, which suggest that the necessary gas conditions
are primarily or exclusively found in GMCs.  Alternatively,
rather than having formed in the field, these stars may have formed in
clusters, and then been subsequently ejected from their birth locations as
runaway stars.  In either case, field massive stars are a unique,
understudied subset of a galaxy's massive star population, probing
both extremely sparse and extremely dense star-forming conditions.

The observational evidence for in situ field massive star formation
has grown in recent years.  An optical and UV photometric census of candidate O-type stars in a portion of the LMC suggests that approximately half of these stars may be members of the field population \citep{Parker01}.  Some strong, direct
evidence of formation in the field is work by \citet{Testi97, Testi98}, who reported a
sample of Herbig Ae/Be stars forming in isolation.  At higher masses,
\citet{Lamb10} detected sparse clusters associated with field OB stars
in the Small Magellanic Cloud, and \citet{Bressert12} identified 15 O
stars that are candidates for isolated formation near 30 Doradus,
based on a variety of criteria.  Additional individual
candidates have been reported by \citet{Selier11} and \citet{Oskinova13}.
\citet{Oey13} presented a sample of 14 field OB stars centered
in circular HII regions, thus implying that they are unlikely to have
transverse runaway velocities.  Since these objects furthermore have
non-runaway radial velocities, they most likely formed in situ.
This growing observational dataset of massive stars that appear to
have formed in sparse clusters or in isolation, without any indication of
being runaways, strongly suggests that some component of the field
massive star population formed in situ.
Even so, formation within clusters cannot be entirely ruled out for these stars.
\citet{Gvarmadze12} point out that cluster dissolution, slow
ejections, or multi-stage ejections could all potentially mask the
signatures that these stars formed in clusters. 

This problem on the origin of field OB stars is central to some
outstanding controversies.  \citet{Weidner06} suggest a deterministic
relation between cluster mass and the associated maximum stellar mass; 
whereas if it is indeed the case that massive stars can form in sparse,
low-mass clusters, it would suggest a large dispersion in the relation
between cluster mass and the associated maximum stellar mass, which is
inconsistent with such a scenario.  Furthermore, it would also imply
that individual sparse clusters must necessarily have stellar initial mass
functions (IMFs) that grossly deviate from any standard values.  It
remains unclear whether such deviations are real or whether they arise from
stochastic sampling, so that an aggregate population of sparse
clusters would yield a Salpeter-like IMF, as suggested by \citet{Lamb10}.

These issues are simply a consequence of
the difficulties in understanding sparse massive star formation
 within the framework of current star formation models.  
Two primary theories for massive star formation are the competitive
accretion model and the core accretion model.  In the competitive
accretion model, molecular clouds fragment into star forming
cores, which continue to accrete matter from a shared reservoir of gas.  In this
scenario, massive stars form in locations where the gas density is
highest, which is typically in the centers of GMCs
\citep{Zinnecker82}.  Thus, it is implicit to the competitive
accretion model that massive stars may only form along with a
significant population of lower mass stars \citep{Bonnell04}.  In
contrast, core accretion models suggest that the gas available for
accretion is controlled by the mass of the fragmented core itself
\citep{Shu87}. Thus in core accretion models it is possible, although
difficult, to obtain gas conditions that would allow a massive star to
form in isolation (e.g. \citealt{Krumholz09}). 

A less controversial component of the field is the runaway population.
Observationally, isolated massive stars with large space velocities
are well-known to exist.  The typical definition for runaway stars is
a peculiar space velocity $> 30$ km s$^{-1}$.  Using this definition,
runaway fractions ranging from $10\%$ \citep{Blaauw61} to $50\%$
\citep{deWit05} have been observed for massive stars within the Galaxy.  However, other studies use evidence from bow shocks, the likelihood of slow runaway
ejections, and the possibility of exotic multi-stage ejection
mechanisms to suggest that the true runaway fraction is much higher,
up to $100\%$ of the field population \citep{Gvarmadze12}.  In this
scenario, the field population is comprised primarily of stars that
formed in dense cluster cores, where the best conditions for massive
star ejections exist.  Thus, the field population is a vital probe of
the massive star formation process at both the densest and least dense
extremes.   

Other than the obvious kinematic signatures expected for runaway
stars, it is not well known how the properties of
massive stars formed in isolation vs runaways would differ from 
stars in clusters.  Observational studies do reveal a few trends:  for
example, a study by
\citet{vandenBergh04} compares the distribution of spectral
types between field and cluster O stars within the magnitude-limited
Galactic O Star Catalog \citep{MaizApellaniz04}, finding that spectral types
for field stars are skewed toward later types than stars in clusters.
Thus, field O stars are either older or less massive as a population
than O stars in clusters.  A similar result was found in the
Magellanic Clouds, where \citet{Massey95} and \citet{Massey02}
discovered that the field population has an extremely steep IMF in a
few selected fields.  The stellar IMF for stars in clusters is generally
consistent with the classical Salpeter slope of $\Gamma = 1.35$ for a
power law given by $dn/ d \log m \propto
m^{-\Gamma}$, where $n$ is the number of stars of mass $m$.
However, \citet{Massey95} found a high-mass field IMF slope of
$\Gamma \sim 4$ using a combination of spectra and photometry.  This
steep IMF also suggests that field massive 
stars are typically less massive than the massive stars in clusters.
These findings represent the largest systematic departure from a
Salpeter IMF based on direct star counts, and suggest that field massive
stars as a population may originate in a fundamentally different way
than those in clusters. 

Thus, there is a clear need for a systematic, statistically complete
survey of field massive 
stars.  Unlike stars in clusters, field massive stars in the
Galaxy are distributed in all directions.  Together with distance
uncertainties and line-of-sight confusion caused by large and
differential extinction, this causes great difficulty in 
identifying a complete, uniformly selected sample of Galactic field O stars.
Sample size and stochasticity are also issues within
the Galaxy, since we are limited to sampling only the nearby Galactic
field.

In order to mitigate these issues, we targeted the nearby Small Magellanic Cloud (SMC) to obtain a uniform, spectroscopic survey of its field massive star population, which we call the Runaways and Isolated O-Type Star Spectroscopic Survey of the SMC, or RIOTS4. Since the SMC is located at high Galactic latitude, it is
relatively free from the line-of-sight issues that plague Galactic
studies.  Additionally, since all objects are at the SMC distance,
RIOTS4 avoids issues associated with distance uncertainties.
Thus, the most important benefit is that RIOTS4 targets a {\it
  spatially complete} and statistically significant sample of
uniformly selected field massive stars.  Here, we present
an overview of the RIOTS4 survey and the results to date.

\section{RIOTS4 Targets and Observations}

RIOTS4 targets a spatially complete sample of 374 uniformly
selected candidate field OB stars in the SMC.  Our targets are 
identified by Oey et al. (2004; hereafter OKP04) according to the photometric criteria
$B\leq 15.21$ and $Q_{UBR} \leq -0.84$, where the reddening-free 
parameter $Q_{UBR}$ is given by,
\begin{eqnarray}
Q_{UBR} &=& (U - R) - \frac {A_U - A_R} {A_B - A_R}(B - R) \nonumber \\
&=&(U - R) - 1.396(B - R)\ ,
\end{eqnarray}
where the $A$ values correspond to extinction in the specified bands.
In the calculation of $Q_{UBR}$, OKP04 adopted the ratio of total
to selective extinction $R_V$ = 3.1 from \citet{Cardelli89}.  These
photometric criteria were designed to select stars with masses 
$\gtrsim 10\ M_\odot$, using the $B$ magnitude to eliminate less massive
main sequence stars, and the $Q_{UBR}$ criterion to identify only the
bluest stars; this corresponds to approximate spectral
types of B0 V, B0.5 I, and earlier.  OKP04 applied these criteria
to the $UBVR$ photometric survey data for the SMC obtained by
\citep{Massey02}, which was optimized to identify OB star candidates.
This survey basically covered the full star-forming 
expanse of the galaxy, which ensures uniform selection of a
spatially complete sample of massive stars in the SMC.
OKP04 further carried out a friends-of-friends analysis on this
sample to identify clusters.  In this algorithm, stars are considered
cluster members if their projected distances to other cluster
members are smaller than the given clustering 
length.  The clustering length is the value that
maximizes the number of identified clusters \citep{Battinelli91},
which is 28 pc for the SMC sample.  Thus the field OB
targets for the RIOTS4 survey correspond to all candidates from
the OKP04 sample with no other candidates within a 28 pc radius.   

OKP04 also identified a sample of candidate field O stars in a
smaller region, covering the SMC bar, using UV photometry from the {\it
  Ultraviolet Imaging Telescope (UIT)} \citep{Parker98}.  These 91 field O star
candidates were selected using reddening-free indices that include UV
and $UBVR$ photometry, along with the same $B$ magnitude criteria as
the main sample.  Of these 91 stars, there are 23 that were not
identified by the optical photometric criteria above.  
We included these stars in our multi-object
observations as described below.

We observed the RIOTS4 survey targets
over a five-year period from 2006 September to 2011 October using
spectrographs on the Magellan telescopes at Las Campanas Observatory.
The majority of our observations were obtained with the
Inamori-Magellan Areal Camera and Spectrograph (IMACS) in the f/4
multi-slit mode on the Magellan Baade telescope \citep{Bigelow03}.
With 49 slit masks, we observed 328 of the 374
candidate field OB stars, or over 7/8 of our total sample.  We also
observed the 23 objects unique to the UV-selected sample
with this setup.  We used the 1200 lines/mm grating and
slit widths of either 0.7$\arcsec$ or 1.0$\arcsec$, yielding
spectral resolutions of  $R\sim$ 3700 and $R\sim$ 2600, respectively.
Due to the varying placement of slits within the slit masks, our
spectral coverage for each star varies; however, every spectrum
includes coverage from 4000 -- 4700 \AA.  We observed each field for a
total of one hour in three exposures of 20 minutes each, which allows
us to achieve a S/N $> 30$ for our fainter targets.  All
observations in our IMACS multi-object campaign occurred between 2006
September to 2010 December.   During our initial observing run in 2006
September one of our 49 fields was observed with the 600 lines/mm
grating, resulting in a spectral resolution of $R\sim$ 1900.

To maximize the multi-object fields, we were unable to include 46 of
our RIOTS4 targets in the IMACS slit masks.  We therefore observed
these targets individually or in pairs using long slit observations. 
The majority of our remaining targets were observed using the Magellan
Inamori Kyocera Echelle (MIKE) Spectrograph on the Magellan Clay
telescope \citep{Bernstein03}.  We also used MIKE to re-observe 29 targets
in cases where important diagnostic
absorption lines fell within the IMACS CCD chip gaps, or when a spectrum from a multi-object observation fell into the center gap of the IMACS CCD array. 
We observed a total of 48 targets with MIKE using a
1$\arcsec$ slit width for a spectral resolution of $R\sim 28000$.
Exposure times for MIKE observations ranged from 15 -- 30 minutes
depending on the brightness of the target, again with a goal of achieving
S/N $> 30$.  All MIKE observations occurred in 2010 November. 
With IMACS f/4 out of commission during our 2011 observations, we also operated IMACS in f/2 mode with a 300
lines/mm grism to observe a total of 27 objects.  Depending on the
seeing, we used either a $0.5\arcsec$ or $0.7\arcsec$ slit width,
which yield spectral resolutions of $R\sim$ 1000 and $R\sim$ 1300,
respectively.  As in the primary IMACS campaign, we observed objects
for a total of one hour, in three 20-minute exposures.  Our IMACS f/2
observations occurred between 2011 July and 2011 October.

We also took advantage of the IMACS multi-object
setup to conduct time-domain monitoring of three of our most densely
populated fields.  As described below in \S \ref{binaryfraction},
our goal was to identify binary stars from radial velocity variations.
We observed about 9 epochs of these fields,
with baselines in time ranging from $< 24$ hours to days, weeks, months,
and years.  Since these fields overlap in area, a few stars were
observed with up to twice as many observations.

Initial reduction of RIOTS4 IMACS multi-slit observations was completed
with the Carnegie Observatories System for MultiObject Spectroscopy
(COSMOS) data reduction package\footnote{COSMOS was written by
  A. Oemler, K. Clardy, D. Kelson, G. Walth, and E. Villanueva.  See
  http://code.obs.carnegiescience.edu/cosmos.}.  COSMOS was custom
designed for use with the IMACS instrument and 8-CCD array setup.
With COSMOS, we performed bias subtraction, flat-fielding, wavelength
calibration, and extraction of 2-D spectra following the standard
COSMOS pipeline.  For single-star spectra from MIKE and IMACS, we used
standard IRAF\footnote{IRAF is distributed by the National Optical
  Astronomy Observatory, which is operated by the Association of
  Universities for Research in Astronomy (AURA), Inc., under
  cooperative agreement with the National Science Foundation (NSF).}
procedures to do bias subtraction, flat fielding, and wavelength
calibration.  From the wavelength-calibrated 2-D spectra for both
single star observations and multi-slit observations, we used the {\tt
  apextract} package in IRAF to find, trace, and extract 1-D spectra.
We rectified the spectra using the {\tt continuum}
procedure and eliminated remaining cosmic rays or bad pixel values with
the {\tt lineclean} procedure, both of which belong to the {\tt
  onedspec} package in IRAF. 

\section{RIOTS4 Data Products}
\subsection{Catalog of Spectral Types}

The first observational data product from RIOTS4 is the catalog of
spectral classifications for candidate field OB stars.  The
completeness of RIOTS4 allows a full characterization of the
distribution of stellar spectral types in the field.
We classify the stars based primarily on the atlas of OB spectra published
by \citet{Walborn90}, and we also rely on \citet{Walborn09} and \citet{Sota11},
especially for identification of unique spectral features.  However,
these atlases 
present mostly Galactic stars at solar metallicity ($Z \sim 0.02$),
which is much higher than the SMC's metallicity ($Z \sim 0.004$).   To
investigate and eliminate potential biases in spectral types due to
metallicity effects, we also refer to \citet{Walborn95} and
\citet{Walborn00} for their comparison of stellar spectral types at
Galactic and SMC metallicity.   To obtain spectral types of supergiant
stars, we adopt the criteria established by \citet{Lennon97} for SMC
metallicity. 

For an initial estimate, four of us (J. B. L., M. S. O., A. S. G., and J. B. G. M.) each
independently estimated the spectral type of every star in the RIOTS4
survey using the above resources.  We collated these spectral types
and arrived at a consensus for each object.  We finalized our catalog by plotting
spectra sequentially from earliest to latest spectral types and
iteratively maneuver stars within this sequence until we achieve a
smooth transition between each spectral sub-type according to
diagnostic stellar absorption line ratios.  The majority of our
spectra are accurate to within half a spectral type, so that, for example, an O8 star can
reasonably be expected to have a spectral type between O7.5 and O8.5.
However, for fainter objects and especially for spectral types later
than B0 V, we sometimes list a range in their spectral types due to
the faintness or non-detection of metal lines caused by a combination of poor S/N and the low
metallicity of the SMC.  Additional difficulties in spectral typing
arise due to confusion from binary systems or Oe/Be stars, which have
emission in one or more Balmer or He lines due to the presence of a circumstellar disk.  
These issues are discussed in more detail below.

\begin{figure*}
	\begin{center}
	\includegraphics[scale=1]{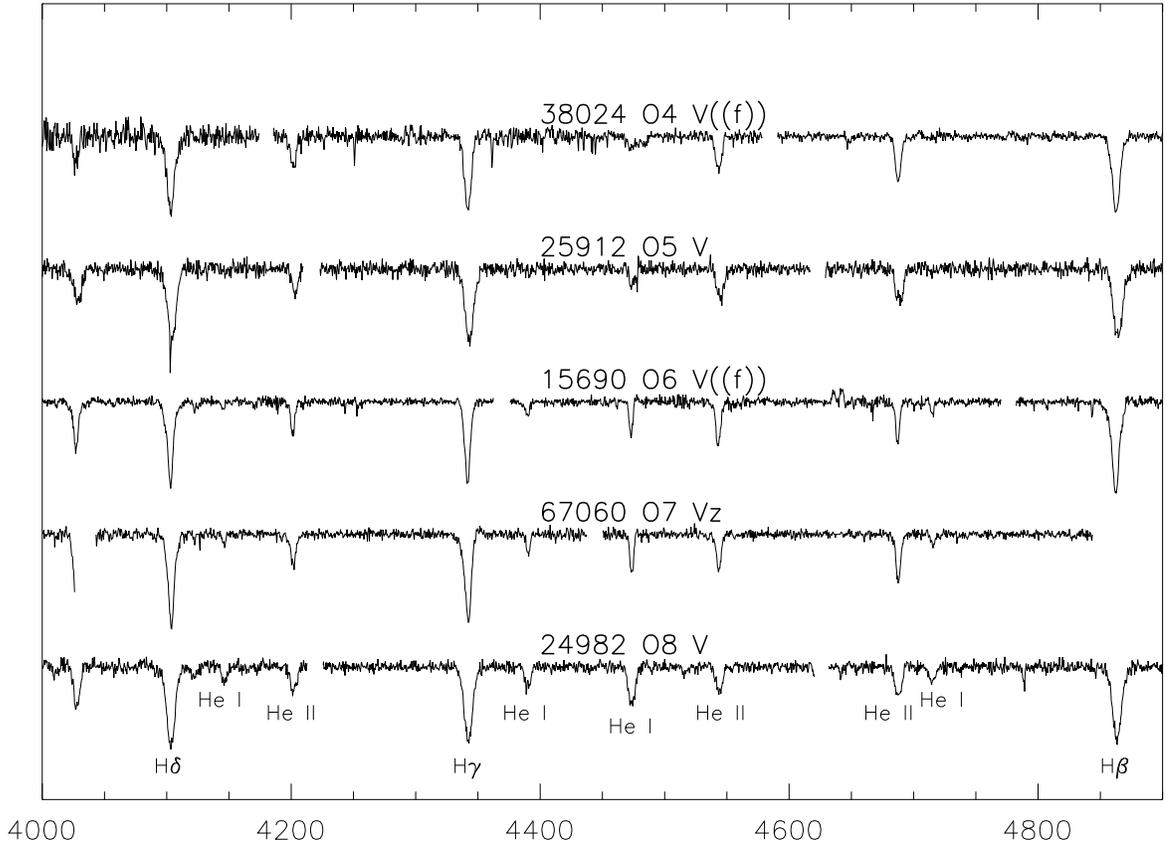}
	\caption{A sequence of spectral types from O4 V to O8 V stars from the RIOTS4 survey.   We label the major spectral features in the range from $4000 - 4900$ \AA.   The ratio of He {\footnotesize II} $\lambda$4542 to He {\footnotesize I} $\lambda$4471 is a primary spectral type diagnostic for O stars.   
	 }
	\label{first}
	\end{center}
\end{figure*}

\begin{figure*}
	\begin{center}
	\includegraphics[scale=1]{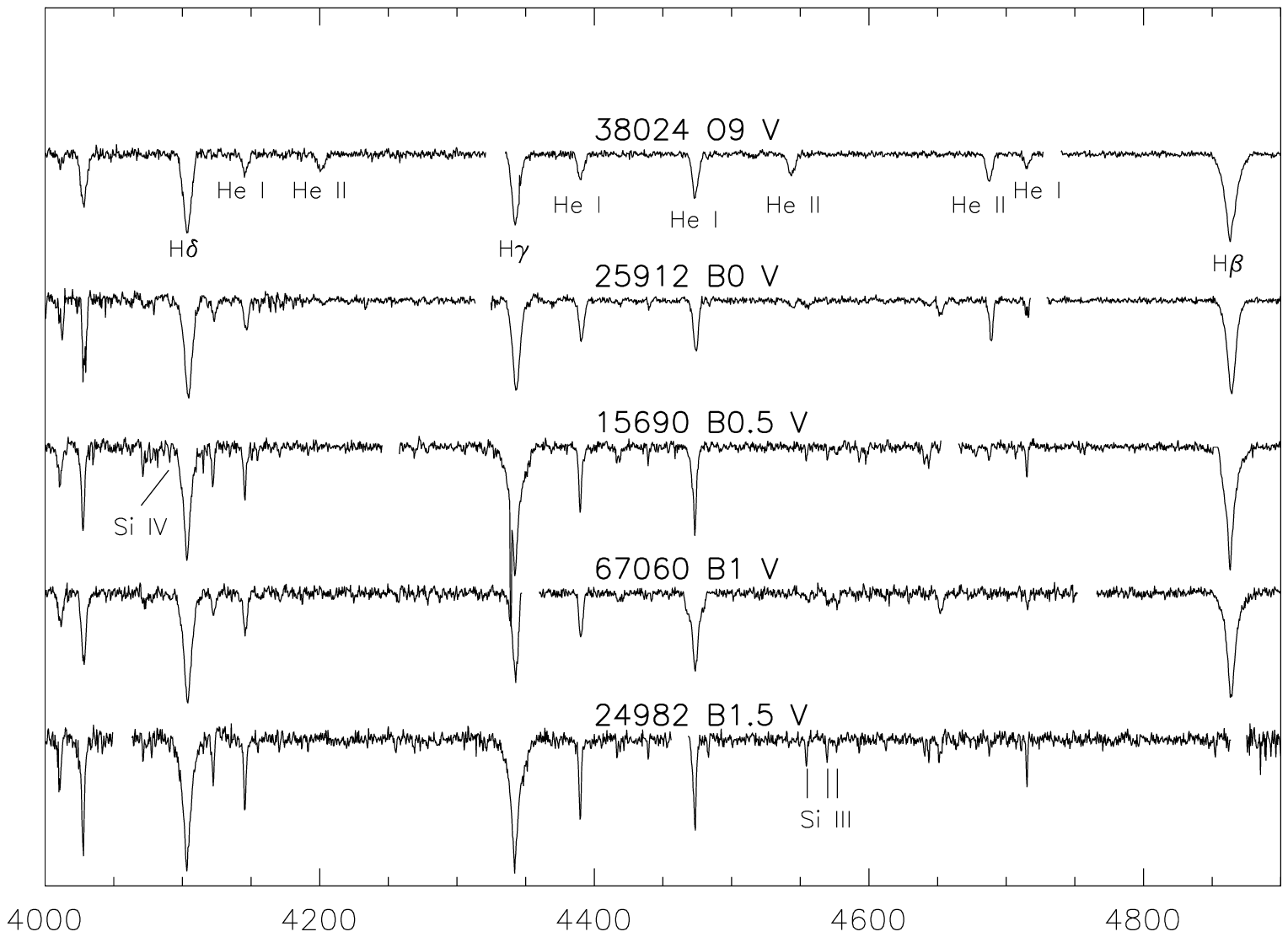}
	\caption{A sequence of spectral types from O9 V to B1.5 V from
          the RIOTS4 survey.   We label the major spectral features in
          the range from $4000 - 4900$ \AA.   With the transition from
          O to B type stars, the primary spectral diagnostic becomes
          the ratio of Si {\footnotesize IV} $\lambda$4089 to Si
          {\footnotesize III} $\lambda$4555, after He {\footnotesize
            II} disappears at spectral type B0.5 V. 
	 }
	\label{last}
	\end{center}
\end{figure*}

We plot a sequence of RIOTS4 spectra in Figures~\ref{first} --
\ref{last}, which cover spectral types from our earliest object, an
O4 V star, to one of our latest objects, a B1.5 V star.
For O stars,
the diagnostic absorption line ratios are He {\footnotesize II}
$\lambda$4542 to He {\footnotesize I} $\lambda$4471 and, as a
secondary check, He {\footnotesize II} $\lambda$4200 to He
{\footnotesize I(+II)} $\lambda$4026.
For B stars,  the primary diagnostic absorption line ratio is Si
{\footnotesize IV} $\lambda$4089 to Si {\footnotesize III}
$\lambda$4555.  A further constraint for early B type stars is the
presence of He {\footnotesize II} $\lambda$4686, which disappears at
spectral types later than B0.2 V, B0.5 III, and B1 I.   

\begin{figure*}
	\begin{center}
	\includegraphics[scale=1]{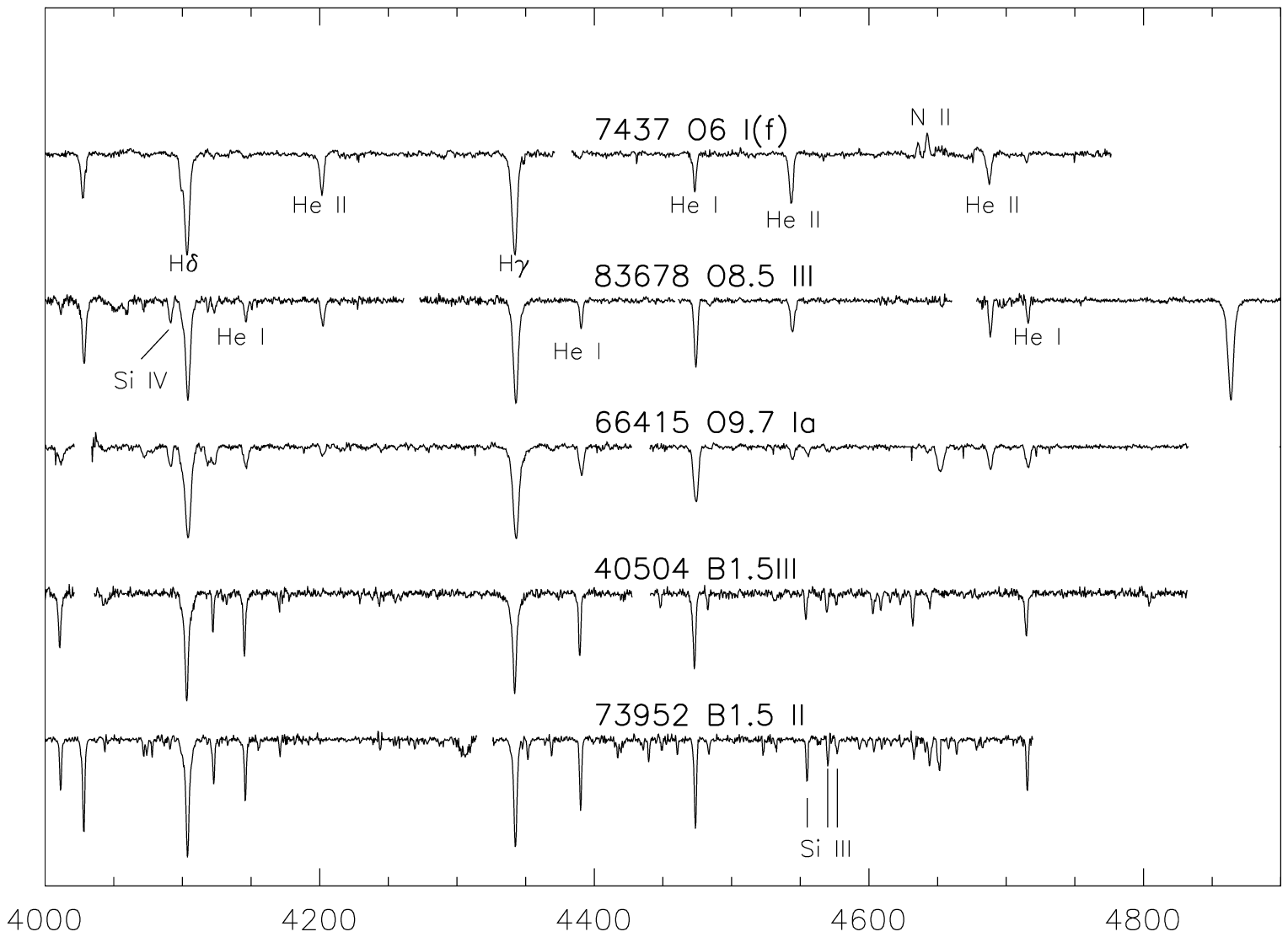}
	\caption{A sequence of evolved stars from O6 to B1.5 from the RIOTS4 survey.  We label the major spectral features in the range from $4000 - 4900$ \AA.  Except for the N {\footnotesize II} emission in the early O stars, evolved luminosity classes are primarily identified by the strength of the Si {\footnotesize IV} for late O stars and Si {\footnotesize III} for early B stars.  
	 }
	\label{evolved}
	\end{center}
\end{figure*}

We determine luminosity classes using a combination of spectral data
as the primary diagnostic, and photometric magnitudes as a secondary
check.  To identify evolved stars at spectral types earlier than
$\sim$O8, we look for the presence of emission features such as N
{\footnotesize II} $\lambda\lambda$4634-4640-4042 and weak absorption
to strong emission in He {\footnotesize II} $\lambda$4686.  For later
O stars, we use the increasing ratio of Si {\footnotesize IV}
$\lambda$4089 to He {\footnotesize I} $\lambda$4026, which identifies
increasingly evolved stars.  In a similar manner, evolution in B stars
is found in the increasing ratio of Si {\footnotesize III}
$\lambda$4555 to He {\footnotesize I} $\lambda$4471.  These luminosity
effects are all demonstrated in the sample of evolved spectra shown in
Figure \ref{evolved}.  As mentioned previously, the lower metallicity
of the SMC causes our spectra to have absent, or much weaker, metal lines than
the Galactic spectral type standards in \citet{Walborn90}.  In
practice, the metal absorption lines tend to be absent in dwarf O
stars for our observational setup, with the exception of {\sc C
  iii} $\lambda$4650 in O9 -- O9.5 V stars, but they do appear in giant
or supergiant luminosity classes.  We
use the spectral criteria for SMC supergiants in \citet{Lennon97} to
finalize our spectral types and luminosity classes for evolved stars.   

For a final check on the luminosity class, we compare the expected
magnitude of our established spectral type \citep{Schmidt-Kaler82} at
the distance of the SMC (DM $= 18.9$; \citealt{Harries03}) with the
observed magnitude from \citet{Massey02}.  If the star is
much brighter than expected for its luminosity class, then we
re-visit our luminosity classification and adjust it to a more evolved
class in more ambiguous cases.  However, the existence of a 
binary companion would also increase the observed brightness of an
object.  Therefore, we carefully re-examine such stars for evidence of 
spectroscopic binary companions.  Even so, detection of a secondary may often go
unnoticed without multi-epoch observations, or they may be
unresolvable due to low inclination angle, small mass ratio, or long
periods.  Thus, undetected binaries may be 
expected to bias our results slightly towards later spectral types and
more evolved objects.  In general, there is a tendency that the
magnitudes indicate brighter luminosity classes than derived
spectroscopically; this is related to the known effect that SMC OB
stars are observed to lie above theoretical evolutionary tracks on the
H-R diagram, as discussed by, e.g., \citet{Lamb13} and
\citet{Massey02}.  However, for Be stars, we find more extreme
discrepancies in luminosity class, and we therefore omit these from
the spectral classifications of Be stars in our catalog.  

\begin{figure*}
	\begin{center}
	\includegraphics[scale=1]{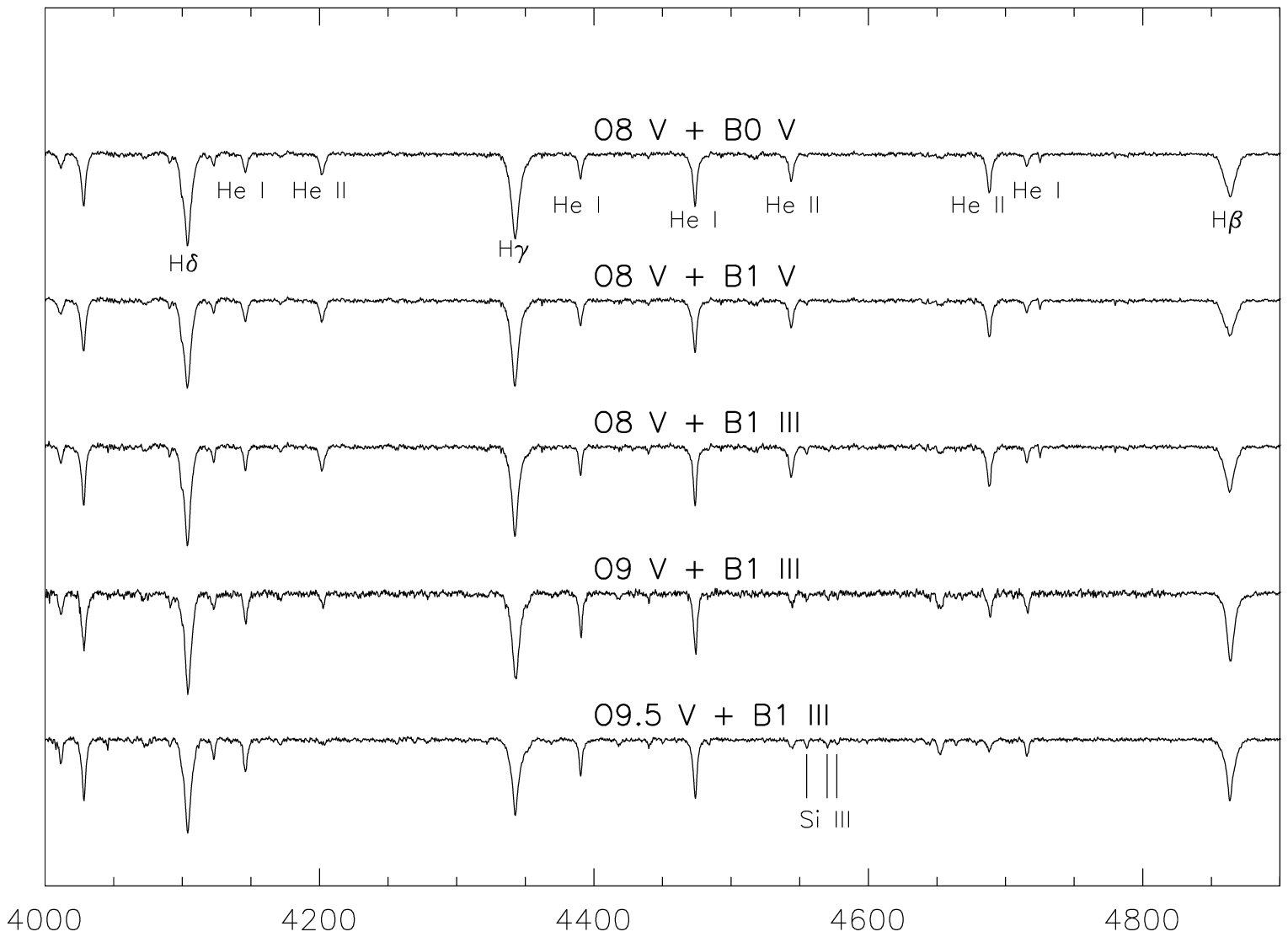}
	\caption{A sample of synthetic binary spectra derived from
          actual RIOTS4 spectra.  We label the major spectral features
          in the range from $4000 - 4900$ \AA.  The top pair of
          spectra demonstrate the difficulty of identifying
          spectroscopic binaries that consist of an O star and a dwarf
          B star, due to the weakness of the Si {\footnotesize III}
          lines.  However, the likely undetected companion does change
          the apparent spectral type of the spectrum from O8 to O8.5.
          In contrast, the bottom three spectra demonstrate the ease
          with which a primary O star can be identified with a giant B
          star companion, due to the clear presence of both He
          {\footnotesize II} and Si {\footnotesize III}. 
	 }
	\label{binarymodels}
	\end{center}
\end{figure*}

The fraction of our objects that are undetected binaries is likely to be
significant; we obtain a lower limit to the binary fraction of $\sim$60\%  in the
RIOTS4 multi-epoch campaign (see \S \ref{binaryfraction}), which is similar
to the frequency found in open clusters
\citep[e.g.,][]{Sana08,Sana09,Sana11}.  Thus, we want to quantify the
potential effects undetected binaries will have on our spectral
catalog.  Furthermore, we require a method to determine spectral types
of identified double-lined spectroscopic binaries.
To address both these concerns, we create a sequence of synthetic
binary stars, which we derive directly from the RIOTS4 spectral data.
We begin by placing RIOTS4 stars with identical spectral classifications into
separate groups.  Any stars that have chip gaps affecting important
diagnostic lines or have poor S/N are removed from these groups.  The
remaining stars in each group are wavelength-shifted to a radial velocity of zero and then median combined to create a template spectrum for each spectral type.  We ensure that each template is
created from a combination of at least five stars, which limits us to
spectral types ranging from O8 to B1.  Using these template spectra,
we combine each pair, weighted according to their expected magnitudes
\citep{Schmidt-Kaler82}, to generate our synthetic binary spectra.  We
plot an example sequence of these synthetic binaries in Figure
\ref{binarymodels}.  From this exercise, we find that the primary star
in the system is rarely altered by more than a single spectral type.
However, we find that the secondary spectral type is poorly
constrained.  This is especially true for O+B binaries, where the
diagnostic Si {\footnotesize III} $\lambda$4555 line for the B star,
which is already weakened due to the low metallicity, is further
affected by the continuum of the primary O star.  Most binary systems
with a B dwarf secondary star are undetectable in RIOTS4 spectra due to the
weak Si {\footnotesize III} lines.

Another stellar population that creates issues for spectral typing is
emission-line stars.  In RIOTS4, this includes classical Oe/Be stars,
supergiant B[e] (sgB[e]) stars \citep{Graus12}, and Wolf-Rayet (WR) stars.  These stars
are often partially or wholly enshrouded in circumstellar disks or
envelopes whose emission is superimposed on the photospheric spectra.
This results in weakened or absent absorption lines, which can
drastically alter spectral types or make them impossible to determine.  
 \citet{Lesh68} classifications for Oe/Be stars were carried out by JBGM.  
In \S \ref{estars}, we summarize our analysis of the four sgB[e] stars from
\citet{Graus12} and present the two WR stars, which are already known in
the literature.  While the number of sgB[e] stars and WR stars
in our sample is small, classical Oe/Be stars account for 42\%
of the RIOTS4 sample (\S \ref{estars}).
Infill of the photospheric absorption lines often affects He {\sc i}
lines in Oe stars \citep[e.g.,][]{Negueruela04}, and Si {\sc iii} $\lambda$4555
and Si {\sc iv} $\lambda$4089 in Be stars.  \citet{GoldenMarx15}
describe in more detail our approach to correcting for this effect in
Oe stars.  

Some stars in the RIOTS4 survey are included in previous spectroscopic
studies of the SMC, including the limited study
of field stars in the Magellanic Clouds by \citet{Massey95} and the
2dF survey of the SMC by \citet{Evans04}.  
Our survey has a typical S/N $\sim 75$ and
$R\sim 3000$, compared to S/N $\sim 75$ and $R\sim 1500$ for
\citet{Massey95}, and S/N $\sim 45$ and $R\sim$ 1600 for
\citet{Evans04}.  A comparison of spectral types for stars in common
with \citet{Massey95} shows agreement to within half a spectral type,
consistent with our internal uncertainty.  The stars in common between the
RIOTS4 and 2dF surveys show similar agreement with spectral type.
However, many stars that we classify as dwarfs in RIOTS4 are listed as giants in 2dF.
This discrepancy is linked to the problematic relation between
observations and theoretical models mentioned above, and appears
to result from our different methods for
determining luminosity classes.  \citet{Evans04} rely more heavily on
stellar magnitude and the equivalent width of H$\gamma$ to determine
luminosity classes due to the relatively poor spectral resolution and
S/N of their data.  Coupling this with the expected
high binary fraction and our careful treatment of
binaries may explain the differences.

\subsection{Stellar Radial Velocities and Multi-Epoch Observations}
Another important RIOTS4 data product is the measurement and
distribution of radial velocities for SMC field OB stars.  Radial
velocities are an important property of a stellar population, both for
individual objects and as an ensemble.  Since runaways are a well
known component of the field population, in principle, we can identify
many such objects
using their radial velocities.  For the field stars as a whole, the velocity
distribution and dispersion probe the kinematics of this population
and on a large scale, the bulk motions of the SMC.  For multi-epoch
observations, variability in the radial velocity is a strong
indicator of a massive binary system.

We measure the radial velocities of RIOTS4 targets using the {\tt
  rvidlines} package in IRAF.  Velocities are obtained by fitting gaussian
profiles to a combination of H, He {\footnotesize I}, and He
{\footnotesize II} absorption lines.  We require a minimum of 3 lines
to determine the radial velocity, to ensure that continuum fitting
issues or odd line profiles do not affect our measurements.  Lines
with velocities that significantly deviate from all other lines for a
single star are excluded from the radial velocity measurement.  These
spurious velocities are typically associated with lines close to the
IMACS chip gaps, which can affect the continuum fitting and,
therefore, the line profile.  The uncertainties on our radial velocity
measurements are $\sim 5$ km s$^{-1}$ for MIKE observations, $\sim 10$
km s$^{-1}$ for IMACS f/4 observations, and $\sim 25$ km s$^{-1}$
for IMACS f/2 observations.   

Since massive stars have a high binary frequency, it is likely that a
large fraction of our radial 
velocity measurements are affected by variability.  Thus, single-epoch
radial velocity measurements may cause erroneous identification of
binary systems as runaway stars.  This variability also adds scatter
to the distribution of radial velocities for the full population.
Our multi-epoch observations are meant to address the magnitude of
these effects by measuring the scatter and estimating the field binary
fraction for 8\% of the RIOTS4 sample (\S \ref{binaryfraction}).  
Our multi-epoch data are all obtained in IMACS
f/4 mode, which gives us sensitivity to radial velocity variations of
$\sim10$ km s$^{-1}$.

\section{Results}

\subsection{Stellar Catalog}
Table \ref{catalog} presents the basic catalog of
the 374 objects in the RIOTS4 survey.  In
columns 1 -- 3, we list the stellar ID numbers and $B,\ V$
magnitudes from \citet{Massey02}, respectively; column 4 contains the
reddening-free $Q_{UBR}$ calculated by \citet{Oey04}.  In
column 5, we provide an extinction estimate using the SMC extinction
maps from \citet{Zaritsky02}.  Column 6 contains the spectral classification
derived from the RIOTS4 data.  Columns 7 and 8 list our measured radial
velocity of the star and the radial velocity of the nearest (in
velocity space) H{\footnotesize I} kinematic component with brightness temperature $>
20$ K (see \S \ref{S_Runaways}).  We list the instrument setup used to
obtain the spectrum in column 9 and the observation date in column
10.  The \citet{Massey02} photometric errors are on average 0.01
at $V = 13.0$, and 0.04 at $V= 15.0$.
Table~\ref{catalogO} provides the same data for the 23 additional
stars we observed from the UV-selected sample.  In what follows, we
consider only the original, optically selected sample so that our
analysis is applied strictly to a uniformly selected sample.  However,
given that there are 23 additional stars out of 91 identified with the
alternate criteria, we can infer that our base sample is incomplete
at least at the 25\% level for identifying all actual OB stars.

\begin{deluxetable*}{cccccccccc}
 \tabletypesize{\small}
  \tablewidth{0pc}
  \tablecaption{RIOTS4 Catalog\tablenotemark{a}} 
  \tablehead{ \colhead{ID\tablenotemark{b}}
    & \colhead{$B$\tablenotemark{b}} & \colhead{$V$\tablenotemark{b}} &
  \colhead{$Q_{UBR}$} & \colhead{$A_V$\tablenotemark{c}} & \colhead{Sp Type} &
  \colhead{RV$_{\rm star}$ }&
  \colhead{RV$_{\rm HI}$\tablenotemark{d}}& \colhead{Instrument} &
  \colhead{Observation Date} \\
  &  &  &  &  &  & \colhead{(km s$^{-1}$)} & \colhead{(km s$^{-1}$)}
& & \colhead{  (YYMMDD)}
}
\startdata
  107 & 14.96 & 15.00 &         -0.95    & 0.82 &   Be$_3$      & -- & -- &    MIKE & 111024 \\
  298 & 15.18 & 15.12 &         -0.91    & 1.03 &   B1e$_{3+}$   & -- & -- &    IMACS f/4 & 070920 \\
 1037 & 15.15 & 15.28 &         -0.85    & 0.44 &   B0.5 V  & 110 & 110 &    IMACS f/4 & 070920 \\
 1600 & 14.42 & 14.60 &         -0.87    & 0.32 &   O8.5 V  &  93 & 103 &  IMACS f/4 & 070920 \\
 1631 & 15.19 & 15.15 &         -0.99    & 1.11 &   B1e$_2$   & 120 & 120 &  IMACS f/4 & 070920
\enddata
\tablenotetext{a}{This table is published in its entirety in the
  electronic edition of the $Astrophysical$ $Journal$.  A portion is
  shown here for guidance regarding its form and content.}
\tablenotetext{b}{From \citet{Massey02}.}
\tablenotetext{c}{From \citet{Zaritsky02}.}
\tablenotetext{d}{Measured from \citet{Stanimirovic99}.}
\label{catalog}
\end{deluxetable*}

\begin{deluxetable*}{cccccccccc}
 \tabletypesize{\small}
  \tablewidth{0pc}
  \tablecaption{Additional UV-Optically Selected Stars in the SMC Bar} 
  \tablehead{ \colhead{ID\tablenotemark{a}}
    & \colhead{$B$\tablenotemark{a}} & \colhead{$V$\tablenotemark{a}} &
  \colhead{$Q_{UBR}$} & \colhead{$A_V$\tablenotemark{b}} & \colhead{Sp Type} &
  \colhead{RV$_{\rm star}$ }&
  \colhead{RV$_{\rm HI}$\tablenotemark{c}}& \colhead{Instrument} &
  \colhead{Observation Date} \\
  &  &  &  &  &  & \colhead{(km s$^{-1}$)} & \colhead{(km s$^{-1}$)}
& & \colhead{  (YYMMDD)}
}
\startdata
 5391 & 13.36 & 13.31 & -1.00 & 1.13 & 	 O8.5 III	&   44 &  98 & 	IMACS f/4 & 060913\tablenotemark{d}  \\
 6946 & 14.60 & 14.69 & -0.87 & 0.57 & 	 O9.5 V	&  141 & 141 & 	IMACS f/4 & 060913\tablenotemark{d} \\
 8257 & 14.69 & 14.49 & -0.87 & 1.55 & 	 B1.5 V	&   96 & 107 & 	IMACS f/4 & 060913\tablenotemark{d} \\
 9534 & 13.63 & 13.76 & -0.84 & 0.44 & 	 B0.2 III	&    - &   - & 	IMACS f/4 & 090824 \\
10129 & 13.87 & 14.01 & -0.87 & 0.45 & 	 B0.2 V	&  130 & 130 & 	IMACS f/4 & 060913\tablenotemark{d} \\
14190 & 15.04 & 14.83 & -0.99 & 1.46 & 	 B1.5 V	&  149 & 149 & 	IMACS f/4 & 090824 \\
15203 & 14.06 & 14.11 & -0.87 & 0.69 & 	 O9.5 V + O9.7 V	&  156 & 156 & 	IMACS f/4 & 060912 \\
15440 & 14.97 & 14.77 & -0.90 & 1.49 & 	 B1e$_3$	&    - &   - & 	IMACS f/4 & 090825\\
15690 & 14.05 & 14.07 & -0.99 & 0.89 & 	 O6 V((f)) 	&   80 & 120 & 	IMACS f/4 & 090824 \\
17963 & 15.12 & 15.21 & -0.99 & 0.55 & 	 B0.2 V	&  115 & 120 & 	IMACS f/4 & 090824 \\
18200 & 14.33 & 14.33 & -0.87 & 0.90 & 	 B0e$_3$	&  111 & 120 & 	IMACS f/4 & 090824 \\
24982 & 14.75 & 14.94 & -0.85 & 0.26 & 	 O8 V	&  110 & 110 & 	IMACS f/4 & 060913\tablenotemark{d} \\
25912 & 14.19 & 14.39 & -0.88 & 0.26 & 	 O5 V	&  150 & 150 & 	IMACS f/4 & 060913\tablenotemark{d} \\
27272 & 13.62 & 13.78 & -0.85 & 0.35 & 	 B0.7 III + B	&  121 & 121 & 	IMACS f/4 & 060913\tablenotemark{d} \\
28153 & 14.69 & 14.83 & -0.88 & 0.41 & 	 O9.5 V	&  169 & 169 & 	IMACS f/4 & 060912 \\
36359 & 14.38 & 14.30 & -1.03 & 1.25 & 	 B1e$_{4+}$	&    - &   - & 	IMACS f/4 & 060912 \\
38302 & 14.64 & 14.81 & -0.84 & 0.29 & 	 B1 V	&  154 & 154 & 	IMACS f/4 & 090825 \\
40341 & 13.77 & 13.98 & -0.92 & 0.24 & 	 O8.5 III((f))	&    - &   - & 	IMACS f/4 & 090825 \\
41095 & 14.84 & 14.85 & -0.92 & 0.92 & 	 O9.5-B0 V + Be$_3$ 	&    - &   - & 	IMACS f/4 & 060911 \\
44634 & 15.19 & 15.37 & -0.85 & 0.27 & 	 O9.5-B0 V	&  150 & 150 & 	IMACS f/4 & 090825 \\
45677 & 13.52 & 13.66 & -0.92 & 0.47 & 	 O9.5 III	&  160 & 164 & 	IMACS f/4 & 090825 \\
48672 & 14.34 & 14.52 & -0.93 & 0.36 & 	 O7.5 V	&    - &   - & 	IMACS f/4 & 090824 \\
53373 & 14.08 & 14.20 & -0.84 & 0.51 & 	 O9 V	&  119 & 122 & 	IMACS f/4 & 090824 

\enddata
\tablenotetext{a}{From \citet{Massey02}.}
\tablenotetext{b}{From \citet{Zaritsky02}.}
\tablenotetext{c}{Measured from \citet{Stanimirovic99}.}
\tablenotetext{d}{Observed multiple times for binary monitoring; see Table 3.}
\label{catalogO}
\end{deluxetable*}

\subsection{Field IMF}
Previous studies of the field massive star IMF in the Magellanic
Clouds indicate a slope steeper than the traditional Salpeter slope of
$\Gamma  = 1.35$.  The observed slopes range from $\Gamma$ = 1.80$\pm 0.09$
\citep{Parker98} to $\Gamma \sim$ 4.0$\pm 0.4$ \citep{Massey95, Massey02}.
However, not all studies agree on this point, as observations of ``field''
stars in the LMC region surrounding 30 Dor suggest an IMF consistent
with Salpeter \citep{Selman11}.
Some of the uncertainty and variation
in these results can be attributed to obtaining the IMF using only
photometry or a combination of photometry and spectroscopy.  As shown
by, e.g., \citet{Massey11}, deriving accurate masses for massive stars can
only be done with spectroscopy.  If spectroscopically determined masses confirm the
steep field IMF then it would represent the largest deviation from the
traditional Salpeter IMF obtained from direct star counts.  RIOTS4 was designed
for such observations, since it avoids the uncertainty of photometric
masses, and our large sample minimizes stochastic effects 
at the highest masses. 

With RIOTS4, we definitively measure the field massive star IMF with our
spatially complete sample of objects; full details on 
methodology and results are reported by \citet{Lamb13}.  Briefly,
for stars with spectroscopically derived masses $> 20\ M_\odot$, we
follow \citet{Koen06} to derive the cumulative mass distribution for
the SMC field and compare it with evolved present-day mass functions
from Monte Carlo models with ages up to 10 Myr, the lifetime of 20
$M_\odot$ stars.  Using this method, we estimate that the
field massive star IMF slope is $\Gamma$=2.3$\pm 0.4$ for the
the highest-mass SMC stars.  This slope is confirmed with OGLE~II
photometry \citep{Udalski98} for $7 - 20\ M_\odot$ stars, using a
stochastic approach that models the uncertainties in stellar
positions on the H-R diagram.
With further Monte Carlo modeling, we determine that
undetected binaries or a unique star formation history are unable to
explain this steep field IMF.  Thus, we conclude that the steep
observed IMF is a real property of the SMC field.  In \S 5, we
attribute this to a preponderance of tiny star-forming events.

\subsection{In Situ Formation of Field O Stars}
As outlined earlier, the origin of the field massive star population is an open question.
In particular, it is unknown whether massive stars are capable of forming
in isolation or within sparse clusters.  Some theories of massive star
formation, such as competitive accretion, suggest that the most massive
star formed in a cluster depends on the cluster mass
\citep{Bonnell04}.  Other theories, such as those based on core
accretion, allow for the formation of massive stars in sparse
environments, or even in isolation \citep[e.g.,][]{Krumholz09}.  The essential
question is whether the formation of massive stars in sparse
environments is merely improbable \citep[e.g.,][]{Elmegreen00} or
actually impossible \citep[e.g.,][]{Weidner06}. 

Using RIOTS4 spectra, along with data from the {\sl Hubble Space
  Telescope (HST)} \citep{Lamb10} and OGLE photometry
\citep{Udalski98}, we identify a sample of unusually strong 
candidates for in-situ, field OB star formation.  \citet{Lamb10} discover three massive
stars that formed in sparse clusters containing $\sim 10$ or fewer
companion stars with mass $> 1M_\odot$ and another three candidates
for truly isolated formation.  \citet{Oey13} present a sample
of 14 field OB stars that are centered on symmetric, dense
{\sc H ii} regions, which minimizes the likelihood that these objects
have transverse runaway velocities.  In both
studies, the RIOTS4 spectra eliminate line-of-sight runaways,
leaving strong candidates for field massive
stars that formed in situ.  We set further constraints on the degree
to which these stars are isolated by examining their immediate stellar
environments with the {\sl HST} and OGLE imaging, allowing us to
evaluate the relationship between the most massive stars in any
sparse clusters and the cluster mass.  Our results imply that these
two quantities are independent, and thus they favor
the core collapse models for massive star formation.

\subsection{Radial Velocity Distribution}
\begin{figure*}
	\begin{center}
	\includegraphics[scale=.45,angle=0]{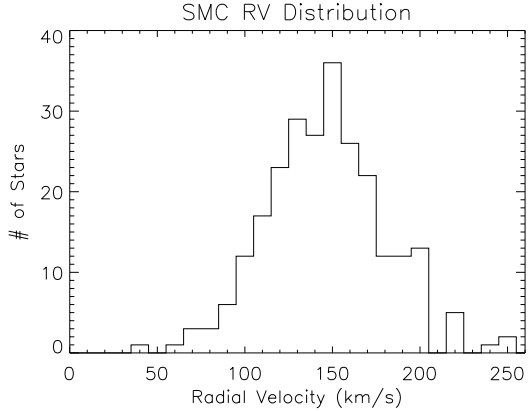}
	\caption{The distribution of radial velocities from stars in the RIOTS4 survey.}
	\label{veldistro}
	\end{center}
\end{figure*}

\begin{figure*}
	\begin{center}
	\includegraphics[scale=.45,angle=0]{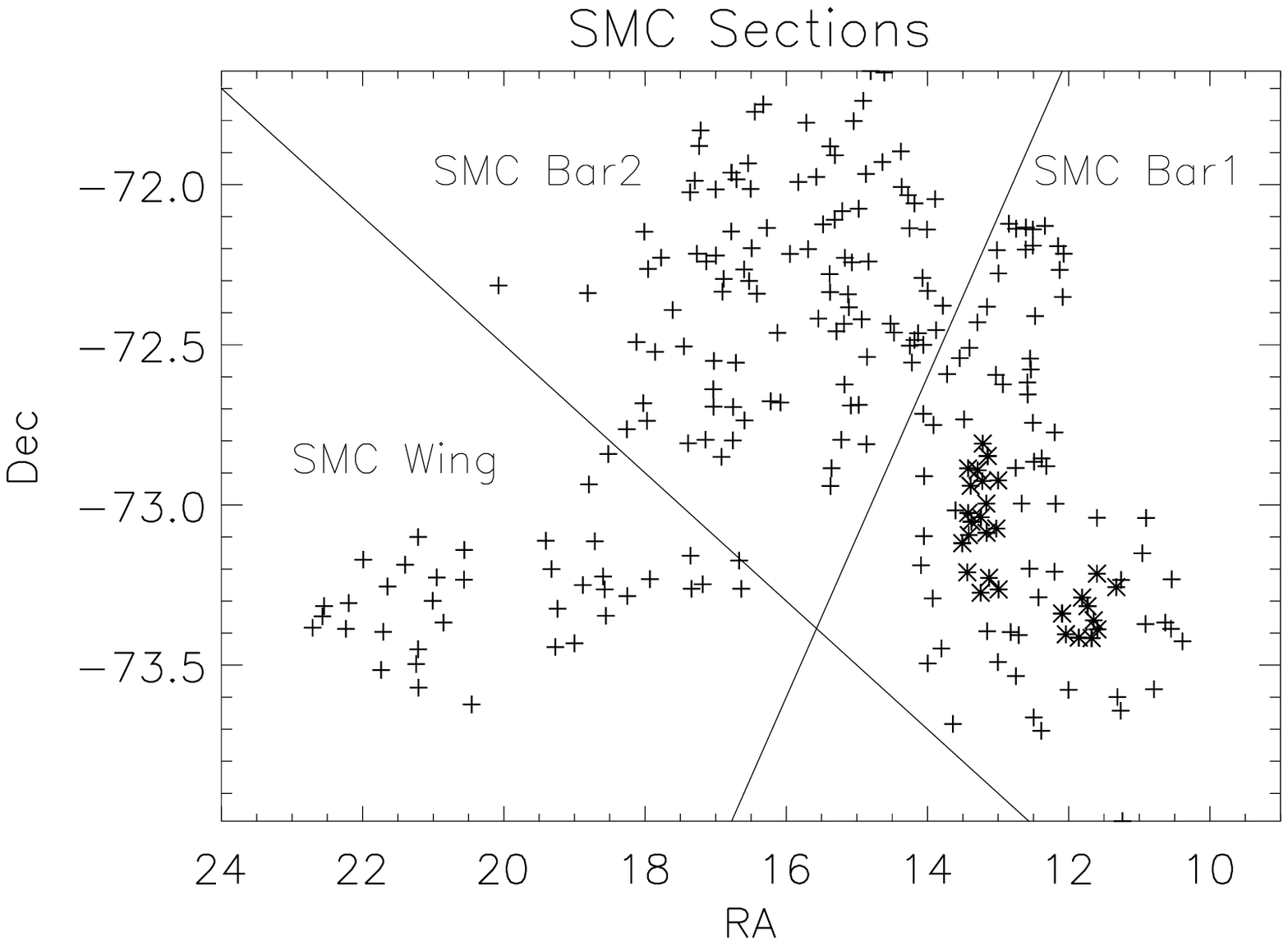}
	\includegraphics[scale=.45,angle=0]{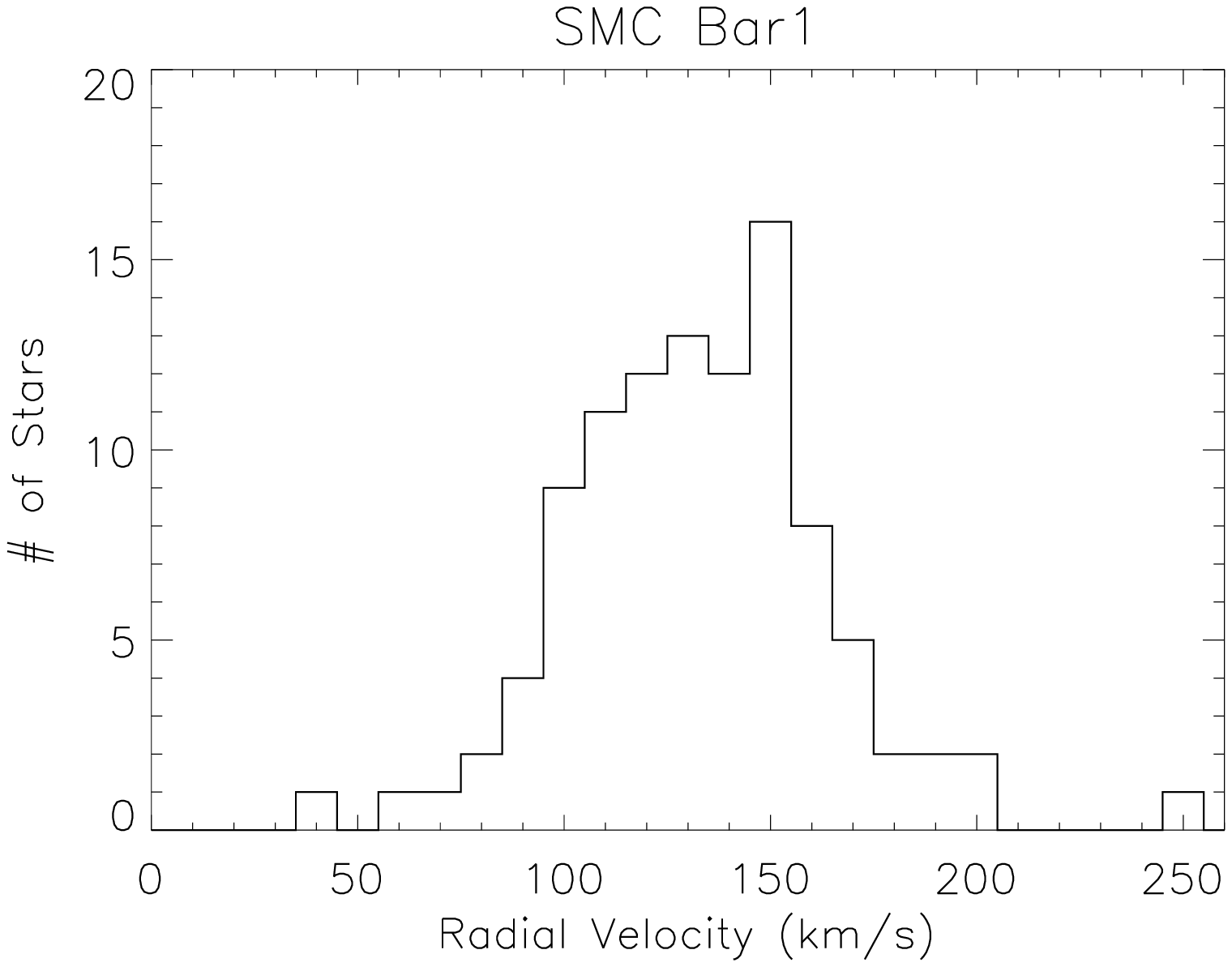}
	\includegraphics[scale=.45,angle=0]{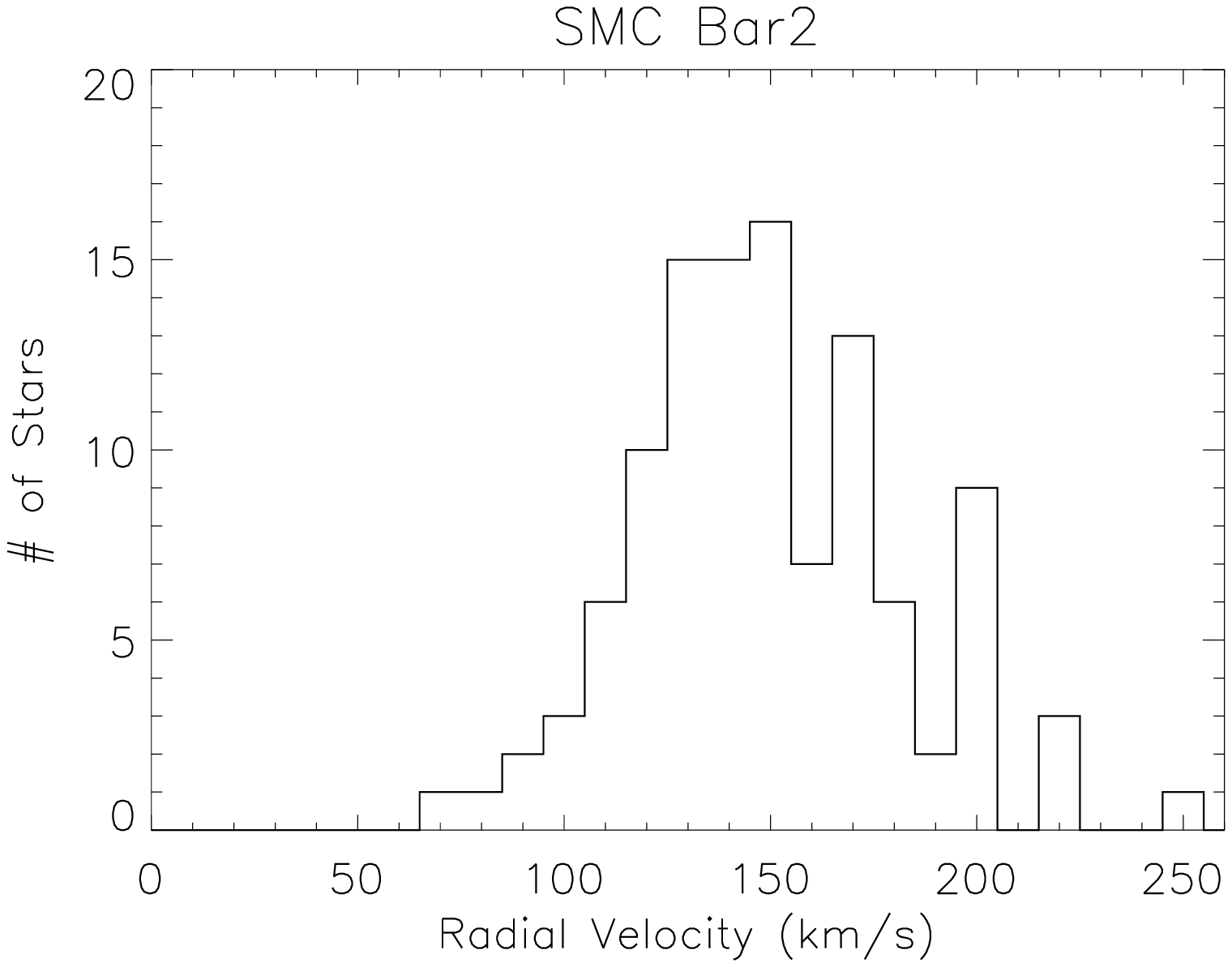}
	\includegraphics[scale=.45,angle=0]{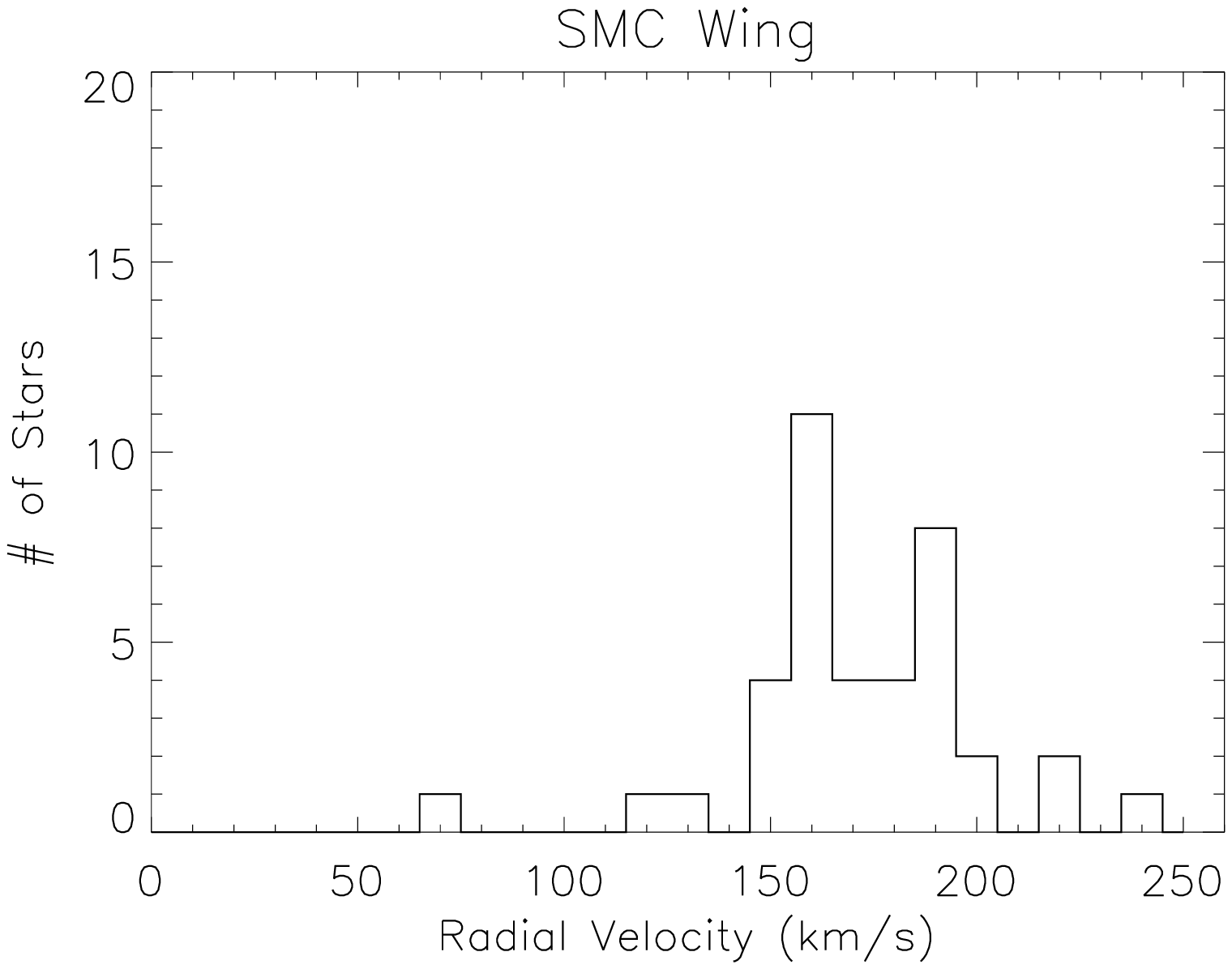}
	\caption{We split the RIOTS4 sample into three regions of the
          SMC, as shown in the upper left panel.  Stars in our binary fields are plotted with asterisks while all other stars are plotted as crosses.  In the other three
          panels, we plot the radial velocity distribution for stars
          in each separate region.  The clear velocity gradient of
          RIOTS4 stars across the SMC agrees qualitatively with the
          velocity gradient of H{\footnotesize I} gas from
          \citet{Stanimirovic04}.} 
	\label{veldistroregions}
	\end{center}
\end{figure*}

The distribution of radial velocities reveals information about the
stellar population kinematics, as well as the bulk motion of the SMC.  
Our velocity distribution from RIOTS4 is generally consistent with
earlier work; Figure \ref{veldistro} is qualitatively
similar to that found from the 2dF survey of OBA-type stars in the SMC
found by \citet{Evans08}.  Both samples exhibit a gaussian-like velocity
distribution with a FWHM of $\sim 30$ km s$^{-1}$ and a mean systemic
velocity of $\sim 150$ km s$^{-1}$.
As mentioned earlier, radial
velocities for individual stars may be affected by binary motions, and
so we can only make inferences based on aggregate trends.
We do see evidence of a
velocity gradient across the SMC, which we depict in Figure
\ref{veldistroregions} by plotting velocity distributions of three
regions in the SMC.  The Bar 1 and Bar 2 regions have mean velocities
of $140$ km s$^{-1}$ and $157$ km s$^{-1}$, respectively, with
corresponding respective velocity dispersions of 32 km s$^{-1}$ and 39 km s$^{-1}$.
Note that although we bisect the bar into two regions, it
appears to have a relatively smooth velocity gradient.  The SMC wing
is more redshifted than the SMC bar, having a mean velocity of $177$ km s$^{-1}$
with velocity dispersion of 29 km s$^{-1}$, but it does not appear to have a
significant internal velocity gradient.  These observations of the
large-scale motions in the SMC agree with results based on stars in the
2dF survey and on {\sc H i} gas from \citet{Stanimirovic04}.

\subsection{Runaway Stars}
\label{S_Runaways}

Runaway stars are a well-known component of the field population, yet
their relative contribution to the field and ejection mechanisms
from clusters remain poorly understood.  Observational estimates for
the runaway frequency range from 10\%
\citep{Blaauw61} to 50\% \citep{deWit05}, while some authors argue that
{\it all} field massive stars are runaways \citep{Gvarmadze12}.  One trend
that seems to have emerged is that O stars have a $2-8$ times
higher runaway frequency than B stars \citep[e.g.,][]{Gies87,Stone91}. 
Runaways arise from one of two likely methods: the binary
supernova scenario \citep{Blaauw61}, or the dynamical ejection
scenario \citep{Poveda67}.  In the binary supernova scenario, the
primary star in a massive binary explodes as a supernova, which drops
the gravitational binding energy of the system and may impart
a kick to the secondary star.  In contrast, dynamical ejections
primarily arise from three- or four-body interactions between a massive
binary and single star or massive binary pairs
\citep[e.g.,][]{Leonard90}.  These ejection mechanisms will imprint
different quantitative properties on the runaway population, including
velocities, binary parameters, and chemical composition.  For example,
the binary supernova scenario cannot produce runaway velocities above
$\sim 200$ km s$^{-1}$, while dynamical ejections can attain
higher velocities \citep[][and references therein]{Gvaramadze09}.
Both ejection scenarios are predicted to include 
binary runaways; however, the type of binaries differ significantly.
For the binary supernova scenario, the compact object remnant of the
primary star sometimes remains bound to the secondary as a runaway
binary system with an eccentric orbit
\citep{McSwain07}.  For dynamical ejections, tight binaries are
sometimes ejected as a single system, thus representing the only
mechanism that can form a runaway double-lined spectroscopic binary.
Finally, while both mechanisms originate from binary systems, stars ejected from
the binary supernova scenario may be He-rich due to contamination from
the supernova explosion \citep{Hoogerwerf01}. 

\begin{figure*}
	\begin{center}
	\includegraphics[scale=.45,angle=0]{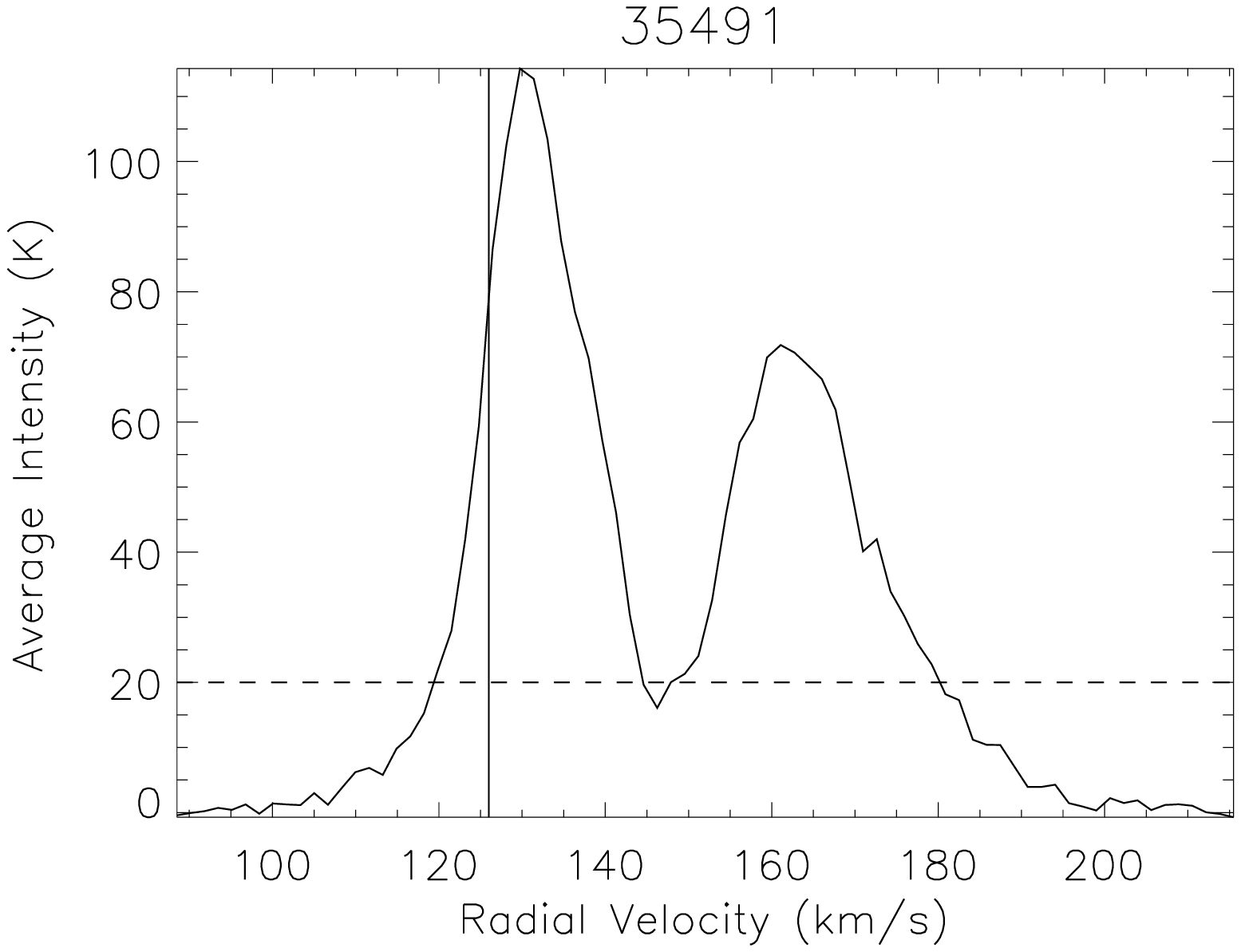}
	\includegraphics[scale=.45,angle=0]{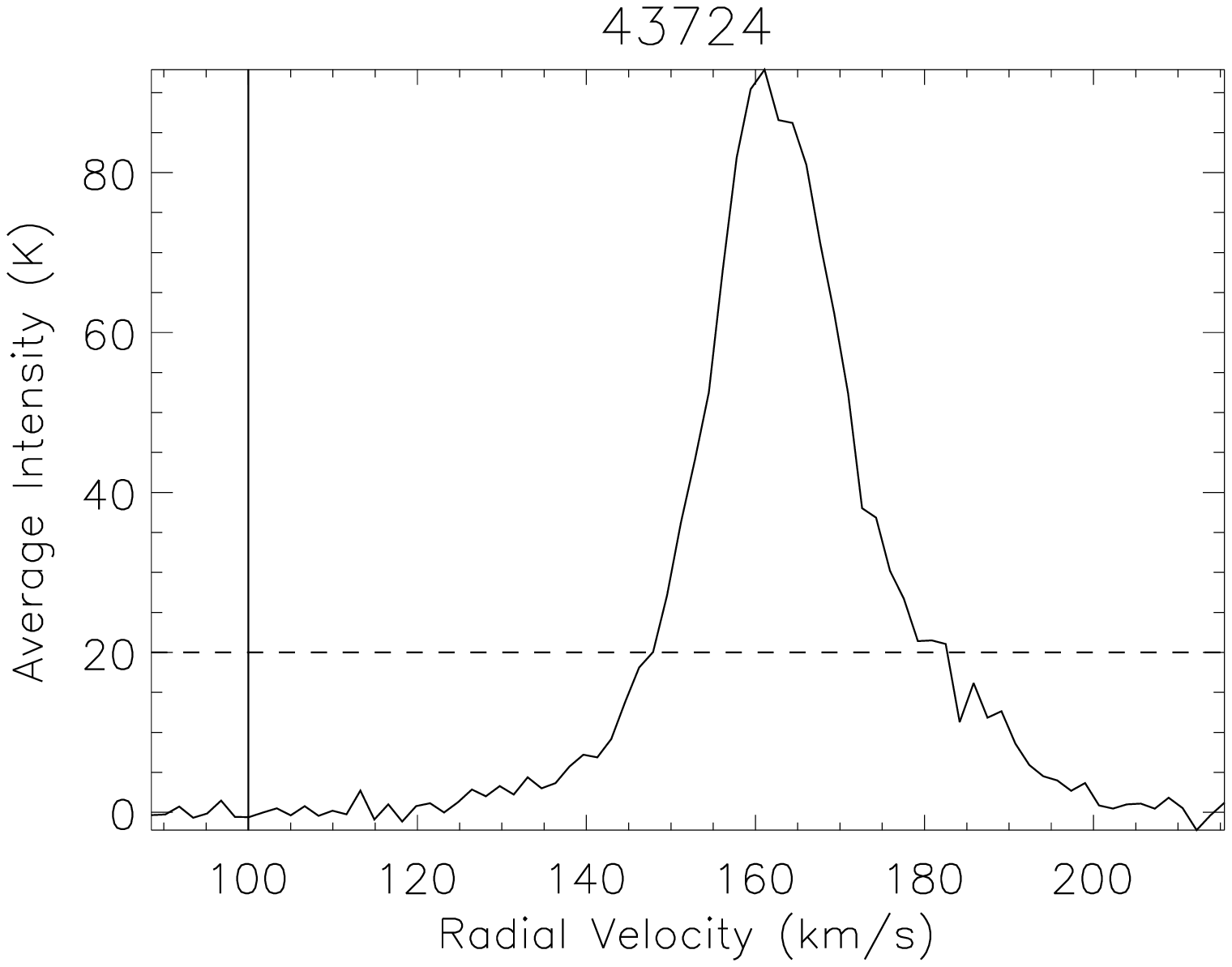}
	\caption{Position-velocity diagrams for H I in the line of
          sight for two RIOTS4 stars, showing data from
          \citet{Stanimirovic04}.  The solid, vertical line depicts the 
          observed radial velocity of the RIOTS4 target, while the
          dashed line shows our brightness temperature threshold of 20
          K.  Stars 35491 (left) and 43724 (right) are examples for which
          the stellar and gas velocities are consistent and
          inconsistent, respectively. }
	\label{runaways}
	\end{center}
\end{figure*}

To estimate the fraction of runaway stars in the RIOTS4 sample, we
compare the observed stellar radial velocities of our OB stars with
the {\sc H i} gas velocity distribution along the line of sight, using
data from the Australia Telescope Compact Array (ATCA) and Parkes
telescopes compiled and mapped by \citet{Stanimirovic99}.  We identify
runaway candidates as those objects with radial velocities that are
different by $> 30$ km s$^{-1}$ from those of the nearest {\sc H i}
velocity components having a brightness temperature $> 20$ K in the
same line of sight.  A pair of examples are shown in Figure
\ref{runaways}, with star 35491 depicting an object consistent with
the line-of-sight {\sc H i} gas velocity, and star 43724
meeting our criteria for a runaway star.  We find that only
11\% of the stars meet these runaway criteria, 27 out of 238 stars
with good radial velocity determinations.
This frequency is likely to be overestimated due to
false positives caused by binary motions, since the
measured radial velocity may reflect the orbital motion for a binary
star, rather than the systemic velocity.  While
such motions will also sometimes cause false negatives
depending on the orbital configuration at the time of observation, 
false positives are more likely to be observed.
A more significant effect is that radial
velocities can only identify line-of-sight runaway motions.  We estimate that
our observations miss 50\% of runaways if the typical ejection velocity is $\sim 60$ km s$^{-1}$.
Since only 8\% of our survey has multi-epoch observations,
we are not yet able to correct for the effect of binaries on the
stellar population kinematics.  Therefore, we have initiated further follow-up,
binary monitoring observations to further minimize these degeneracies. 

We do find one runaway, star 5391, that we identify as a binary star 
from our multi-epoch observations (\S~\ref{binaryfraction}).  Its
radial velocity of 44 km s$^{-1}$ is 55 km s$^{-1}$  
removed from the nearest significant component of {\sc H i} gas.  With
a semi-amplitude of 108 km s$^{-1}$ in our observed variations, the secondary 
cannot be a degenerate star.  Therefore, if this binary system is indeed ejected from
a cluster, then it must be due to the dynamical ejection mechanism
rather than the binary supernova mechanism.  Since dynamical ejection
frequently splits binaries, the existence of a non-degenerate, runaway
binary suggests a major contribution by this process to the runaway population.

Another interesting object also points to the importance of dynamical
ejection:  Star 49937 appears to be an extreme runaway that is unlikely to
be the product of the binary supernova scenario.  Its
runaway velocity is $\sim 200$ km s$^{-1}$ removed from the nearest
{\sc H i} velocity component.  While it is possible that
this star's runaway component is completely in the line of sight
and/or fortuitously enhanced by binary motion, its high radial
velocity, taken at face value, is near the maximum ejection speed
possible from the binary supernova mechanism \citep{PortegiesZwart00},
as mentioned above.  Thus, the existence of this star again suggests a
significant role for dynamical ejection of runaways.

\subsection{Binary Stars}
\label{binaryfraction}
Stellar multiplicity is a key parameter that probes the formation and
dynamical evolution of a stellar population.  For example,
large protostellar disks may be disrupted in high-density
environments, thereby suppressing the formation of massive binaries
\citep{Kratter08, Kratter10}.
Recent studies of Galactic clusters and OB associations find observed
massive-star binary fractions ranging from $\sim 60\%$ to $\sim 80\%$ 
\citep[e.g.,][]{Sana08,Sana09,Sana11,Kiminki12,Kobulnicky14}.
However, few studies have systematically investigated the multiplicity
of massive stars in the field.  Early studies found that field massive
stars have roughly half the binary frequency of massive stars found in
clusters \citep[e.g.,][]{Stone81,Gies87}.  This general trend of a
lower field binary frequency persists in later studies, such as
\citet{Mason98,Mason09}, who use speckle interferometry of objects in
the Galactic O Star Catalog \citep{MaizApellaniz04} to compare the
frequency of multiplicity between 
cluster and field O stars.  In this magnitude-limited
sample, they find a $39\%$ binary fraction for field O stars, compared
to a $66\%$ binary fraction for O stars in clusters.  When combining
their results with data from the literature on spectroscopically
identified binaries, they obtain 51\% and 75\% binary fractions for field
and cluster O stars, respectively.  However, the spectroscopic data
for these objects is non-uniform and therefore may not provide an
accurate comparison of these statistics between cluster and field O
stars.  But it does suggest that the frequency of multiplicity
for massive stars in the Galactic field is lower than in clusters. 

With the RIOTS4 survey, we performed repeat observations of three
IMACS slit-mask fields over the 5-year survey period,
totaling 29 objects, to obtain an initial evaluation of the binary fraction of field
massive stars in the SMC.  We note that some of these stars belong to
the UV-selected sample (Table~\ref{catalogO}), rather than the default
sample.  We have $9-10$ epochs for each field, at
intervals of days, weeks, months, and years apart; three stars
appear in two of the three fields, yielding up to twice the number of
observations for these objects.  As with the larger survey, these
fields have a high fraction of Oe/Be stars, and we focus here
primarily on the 17 non-Oe/Be stars in these fields.  We use three
separate methods to identify potential binaries, which are described
below.  The first method identifies binaries using 
maximum observed radial velocity variations, the second method is based
on a statistical F-test analysis following \citet{Duquennoy91}, and
the third method uses the period power spectrum and searches for
binary orbital solutions from the radial 
velocity data.  Table~\ref{t_binaries} summarizes this binary
monitoring sample:  columns 1 and 2 give the star ID and spectral
type, respectively; column 3 gives the number of observations, and
columns 4 and 5 show the star's binary status determined from the
second and third methods; we note that the first method yields the
same results as the second.  Column 6 gives the systemic velocity
based on the orbital solution, if available, or the mean of the
minimum and maximum measured radial velocities.  Column 7 gives the
largest velocity variation observed within a 14-day interval $\Delta
v$, and column 8 provides the standard deviation $\sigma_{\rm obs}$ of
the radial velocity measurements for each star.  Column 9 lists the
calculated $P (\chi^2)$, which is 
used to determine binary status in the statistical F-test (\S 4.6.2). 
Column 10 shows the
observation dates for each object, coded as indicated.

\begin{deluxetable*}{llcccccccc}
 \tabletypesize{\small}
  \tablewidth{0pc}
  \tablecaption{Stars in binary monitoring fields}
\tablehead{ \colhead{ID} & \colhead{SpT} & \colhead{$N$}
  & \colhead{F-test} & \colhead{Power Spec} & $v_{\rm sys}$ (km s$^{-1}$) & 
  $\Delta v$ (km s$^{-1}$) & $\sigma_{\rm obs}$ (km s$^{-1}$) & $P (\chi^2)$
  & \colhead{Observation Dates\tablenotemark{a}}
}
\startdata
\multicolumn{6}{c}{Normal OB Stars} \\
\hline
5391  & O8.5 III  &  9  & Y  & Y & 44 & 144  & 75  & $<0.01$  &  ABCEFGIJK  \\
6908  & O9.5 -- B0 III &  9  & Y & N &  128  & 93  & 25  & $<0.01$  & ABCEFGIJK  \\
6946  & O9.5 V &  9 & N & N & 141 & 35  & 12  & 0.74  &   ABCEFGIJK  \\
7437  & O6.5 I(f)    &  9  & Y  & Y & 151 & 29  & 33  & $<0.01$  &   ABCEFGIJK  \\
7782  & O8 V   &  9  & Y & Y & 127 & 65  & 33  & $<0.01$  &   ABCEFGIJK  \\
8257  & B1.5 V  &  9  & Y & Y & 96 & 61  & 21  & $<0.01$  &   ABCEFGIJK  \\
8609  & B0 III  &  9  & N & N & 128 & 21  & 11  & 0.97  &   ABCEFGIJK  \\
10129  & B0.2 V  &  9  & Y & Y & 130\tablenotemark{b}  & 29  & 21  & $<0.01$  &   ABCEFGIJK \\
10671  & B0.5 V    &  9  & Y & N & 122 & 108  & 33  & $<0.01$  &   ABCEFGIJK  \\
21844  & O8 III((f)) &  9  & N  & N & 151  & 36  & 13  & 0.09  &   BCDEFGHIK \\
24213  & B0 III   &  16  & N & N & 126  & 6  & 9  & 0.99  &   ABCDEFGHIJK\\
24982  & O8 V  &  8  & Y & Y & 110  & 59  & 32  & $<0.01$  &   ADFGHIJK \\
25912  & O5 V  &  9  & Y & Y & 150  & 103  & 45  & $<0.01$  &   ADEFGHIJK \\
27272  & B0.7 III + B &  9  & Y  & Y & 121\tablenotemark{c}    & 223  & 105  & $<0.01$   & ADEFGHIJK \\
27600  & B0.5 III  &  10  & N & Y & 177\tablenotemark{b}  & 16  & 13  & 0.64  &   BCDEFGHIJK \\
27712  & B1.5 V    &  7  & N & N & 127  & 8  & 7  & 0.46  &   ADFGHJK \\
28841  & B1 III   &  10  & N & Y & 141  & 22  & 15  & 0.02  &   BCDEFGHIJK \\
\hline \\
\multicolumn{6}{c}{Classical Oe/Be Stars} \\
\hline
7254  & O9.5 IIIe$_2$   &  9 &\nodata & Y & 126  & 10  & \nodata & \nodata & ABCEFGIJK \\
21933  & Be$_3$    &  4 &\nodata & N & 130   & 57  & \nodata & \nodata & AHIJ \\
22321  & O9.5 IIIpe$_{4+}$  &  10 &\nodata & Y & 167\tablenotemark{b}   & 28  & \nodata & \nodata & BCDEFGHIJK \\
23710  & O9--B0 pe$_{3+}$ &  10 &\nodata & N & 168   & 48  & \nodata & \nodata & BCDEFGHIJK \\
23954  & B1.5e$_{3+}$  &  7 &\nodata & N & 130   & 69  & \nodata & \nodata & ADFGHIJ \\
24229  & B1e$_2$    &  7 &\nodata & N & 155   & 19  & \nodata & \nodata & ADFGHJK \\
24914  & O9 III-Vpe$_1$    &  4 &\nodata & Y & 81   & 20  & \nodata & \nodata & AEHI \\
25282  & B0e$_1$   &  17 &\nodata &  N & 130   & 72  & \nodata & \nodata & ABCDEFGHIJK \\
25337  & Be$_3$   &  9 &\nodata & Y & 124   & 55  & \nodata & \nodata & BCDEFGIJK \\
27135  & B1e$_2$    &  18 &\nodata &  N & 113   & 30  & \nodata & \nodata & BCDEFGHIJK \\
27736  & B0e$_2$   &  6 &\nodata & Y & 153   & 39  & \nodata & \nodata & DEFGHJ \\
27884  & O7-8.5 Vpe$_{4+}$ &  10 &\nodata & Y & 156  & 32  & \nodata & \nodata & BCDEFGHIJK 
\enddata
\tablenotetext{a}{Dates of observation are coded as follows: (A) 2006 September 13, (B) 2007 September 19, (C) 2007 September 20, (D) 2008 September 24, (E) 2008 October 6, (F) 2008 October 7, (G) 2008 October 11, (H) 2008 November 21, (I) 2008 November 22, (J) 2009 August 24, (K) 2010 December 20.}
\tablenotetext{b}{From orbital solution.}
\tablenotetext{c}{Average of SB2 components A and B.}
\label{t_binaries}
\end{deluxetable*}

\subsubsection{Maximum radial velocity variation and timescale}

\begin{figure*}
	\begin{center}
	\includegraphics[scale=.45,angle=0]{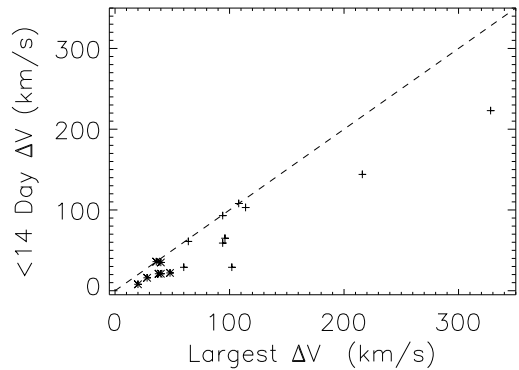}
	\caption{The observed maximum short-term ($< 14$
          days) radial velocity difference vs the largest 
          radial velocity difference over any period.  The dashed line
          depicts the identity relationship. Objects with the highest observed velocity 
          difference happening over a $< 14$ day period will lie on this line.  We expect real binary systems will exhibit velocity variations on both long and short term periods.
          Binaries identified by having
          radial velocity variations $> 30$ km s$^{-1}$ are plotted with a
          plus sign, while single stars are depicted as asterisks.} 
	\label{ampvel}
	\end{center}
\end{figure*}

To identify binary star candidates, we first compare the
amplitude of radial velocity variations with the timescale of the
variations.  Since the amplitude of radial velocity variations is
inversely correlated with the period of a binary system, binaries with
large-amplitude variation should display variability on short
timescales, provided the eccentricity of the system is near zero.  In Figure \ref{ampvel}, we plot the amplitude of the
maximum observed radial velocity variation over short timescales ($ <
14$ days; Table~\ref{t_binaries}) versus the amplitude of the maximum radial velocity
variation over any time scale.  Note that in Figure \ref{ampvel} all
objects must lie at or below the dashed identity line.  In an ideal
scenario, all short-period systems will lie along this locus;
however, we cannot expect good sampling with $\lesssim 10$
epochs of data.  Nonetheless, we still observe a large fraction of
high-variation systems along the identity line, which suggests there
are no systematic velocity offsets over time.  Given the sampling of these fields
and our systematic errors, we conservatively identify binaries as those
objects with radial velocity variations $> 30$ km s$^{-1}$ including
errors.  This yields 10 probable binaries out of the
17 non-Oe/Be stars in our binary monitoring fields.   

\subsubsection{F-test:  radial velocity variations relative to noise}

\begin{figure*}
	\begin{center}
	\includegraphics[scale=.45,angle=0]{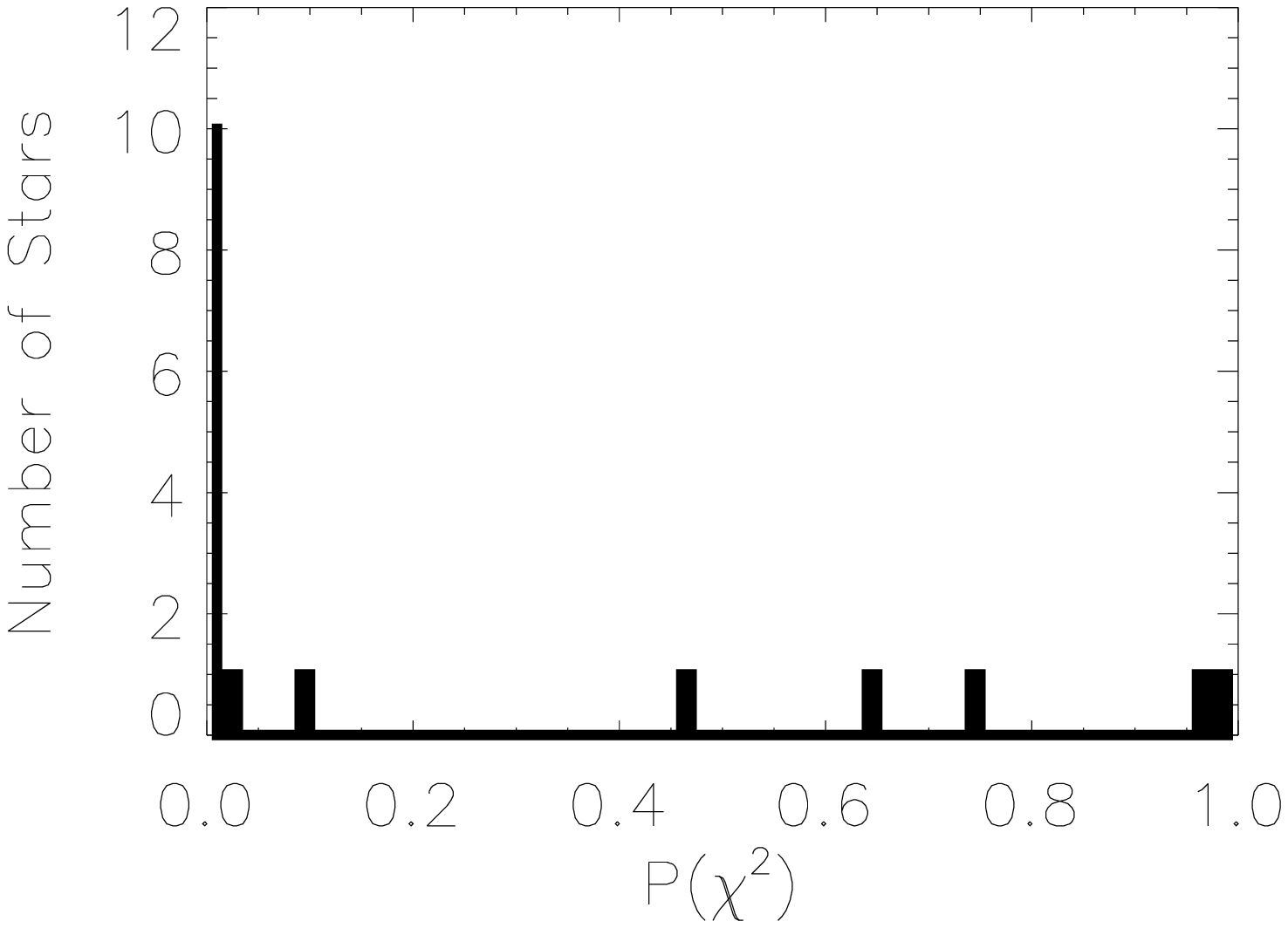}
	\caption{The distribution of $P (\chi^2)$ for non-Oe/Be stars in
          our binary fields.  
          Binary systems that exhibit radial velocity variations
          significantly larger than expected from observational errors
          will have very low $P (\chi^2)$ values ($< 0.01$). }
	\label{ftest}
	\end{center}
\end{figure*}

We also use the approach of
\citet{Duquennoy91} who identified binary candidates in the
nearby solar neighborhood.  This method compares the mean of the
statistical measurement errors associated with each radial velocity measurement
($\sigma_{\rm ave}$) with the standard deviation in the measured radial
velocities ($\sigma_{\rm obs}$; Table~\ref{t_binaries}) for each star.  For single objects with
properly estimated measurement errors, the ratio of
$\sigma_{\rm obs}$/$\sigma_{\rm ave}$ should approximately equal unity.
However, it is unclear where the cutoff ratio between single objects
and binary stars should occur.  Thus, \citet{Duquennoy91} use a
statistical F-test to measure the probability $P (\chi^2)$ that the
observed variations are due to statistical noise.  Following 
their work, we calculate $\chi^2$, accounting for the number of
observations $n$, with:
\begin{equation}
\chi^2 = (n-1)(\sigma_{\rm obs}/\sigma_{\rm ave})^2 \quad .
\end{equation}
Using the cumulative chi-square distribution given by
\begin{equation}
F_k(\chi^2) = G (k/2 , \chi^2/2)
\end{equation}
where $G$ is the regularized Gamma function for a given degree of
freedom $k = n-1$, we calculate $P (\chi^2) = 1-F_k(\chi^2)$, given in
Table~\ref{t_binaries}. 
In the case that all objects are single, the distribution of $P(\chi^2)$ 
should be uniform between values of 0 and 1.  Binary systems, on the
other hand, should have very low values of $P(\chi^2)$, since
their radial velocity variations are not due to statistical noise.
Thus, we can identify binaries as those objects with $P (\chi^2) < 0.01$.
We plot the distribution of $P (\chi^2)$ for the same 17 stars in
Figure \ref{ftest}.  Again, we find a high binary fraction with 10 out
of 17 objects having $P (\chi^2) < 0.01$. 

\subsubsection{Period power spectrum}

\begin{figure*}
  \plottwo{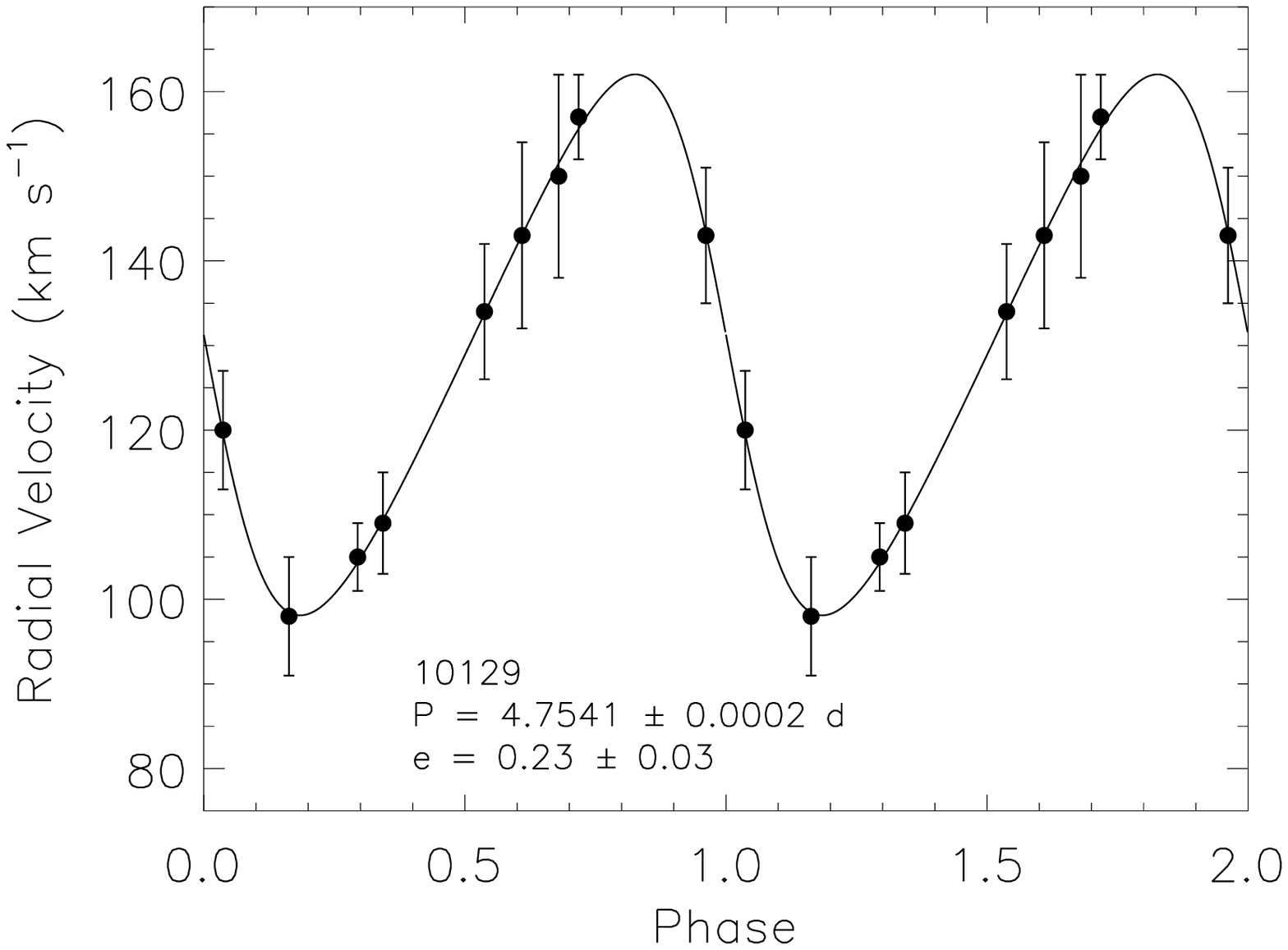}{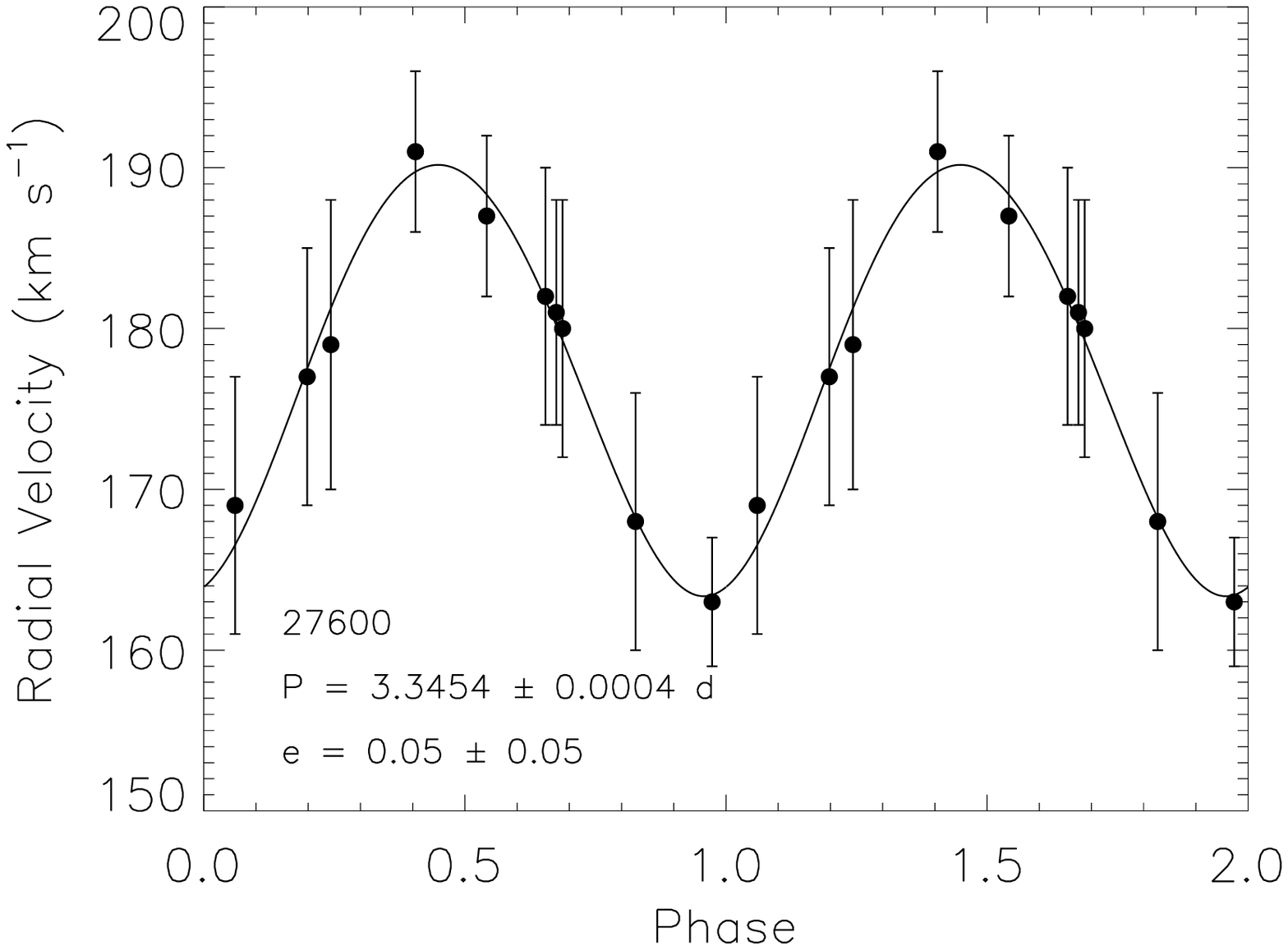}
\caption{Phase diagrams showing the solutions for two more
          securely identified binaries among the normal OB stars in
          the monitoring sample. \label{f_orbits}
}
\end{figure*}

We used the radial velocities to search for credible orbital solutions for all
the stars in the binary monitoring sample based on the method
described by \citet{Kiminki12a}.  In this approach, we generate the
power spectrum of periods for each object, and identify the most
likely values, if any, with an IDL program
created by A. W. Fullerton, which uses the CLEAN deconvolution algorithm of 
\citet{Roberts87}.
We then apply the \citet{gudehus01} code for
determining orbital solutions, the Binary Star Combined
Solution Package, using the candidate periods.  We show the phase
diagrams of the two best orbital solutions in Figure~\ref{f_orbits}.
These are for stars 10129 and 27600, with periods around 4.8 and 3.3
days, respectively.  Star 10129 appears to have a moderate
eccentricity around $e = 0.2$, while 27600 is consistent with a purely
circular orbit.  This approach again
yields 10 out of 17 probable binaries, although the identified
candidate binaries are not the exact same ones found with the
preceding methods (Table~\ref{t_binaries}).

\subsubsection{Binary Fraction}

All three binary identification methods suggest binarity in
10 out of 17 ($59\% \pm 12\%$) of the non-Oe/Be stars in our three binary
monitoring fields.  This frequency is consistent, within the uncertainty,
with previous observations of multiplicity in the Galactic field,
which are $\sim 40 - 50$\% as described above.  However, the small number
statistics generate large errors, and our binary frequency is actually
closer to values observed in Galactic clusters and OB associations. 
It is further difficult to compare these frequencies because of
the different observational biases inherent in the different binary
detection methods and sample properties; our frequencies are
lower limits, representing results only for spectroscopic binaries.
\citet{Sota14} find a strong lower limit of 65\% for the combined
spectroscopic and visual binaries in the southern component of their
Galactic O star survey.  Almost one quarter of these are identified
exclusively by astrometric methods.  We have started follow-up
monitoring of additional RIOTS4 targets
to confirm these results, and to obtain binary orbital parameters.

We also applied the third binary identification method to the
remaining 12 stars in the monitoring fields, which are classical Oe/Be
stars.  The radial velocities measured for these stars are more
uncertain than for normal stars because of emission-line contamination
in the H lines.  We find that 6 out of the 12 Oe/Be stars appear to be probable
binaries.  

\begin{figure*}
	\begin{center}
	\includegraphics[scale=1]{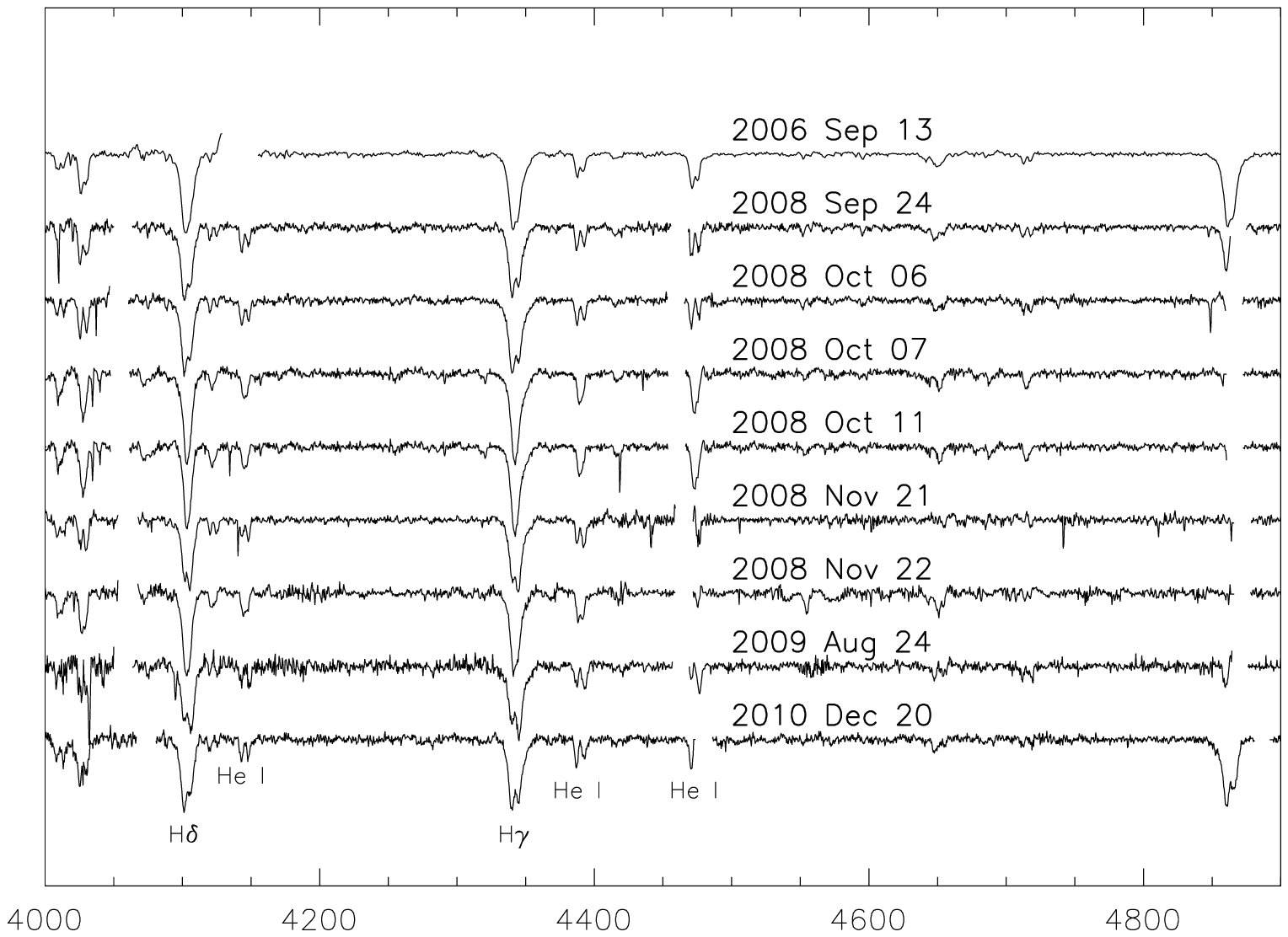}
	\caption{The multi-epoch, RIOTS4 spectra of the double-lined
          spectroscopic binary 27272, with observation dates shown.
	 }
	\label{27272}
	\end{center}
\end{figure*}

One of our binaries, star 27272, is a double-lined spectroscopic
binary (SB2) with B0.7~III and B star components (Figure \ref{27272}).  In our
observations of this system, we find that the stronger absorption line
appears blueshifted in all but 1--2 epochs.  While this may be evidence
of the Struve-Sahade (S-S) effect \citep{Struve37,Sahade59}, 
it is most likely caused by an unfortunate observing cadence,
which impedes our ability to obtain a satisfactory orbital solution.

\subsubsection{Systemic Velocities}

Estimated systemic velocities $v_{\rm sys}$ for the 29 stars in the
monitoring fields are given in Table~\ref{t_binaries}.  These are
generally given as the average of the minimum and maximum of the
$N$ radial velocity measurements for each star.  For three objects,
more reliable values are available from fitted orbital parameters.  
The mean $v_{\rm sys}$ is 131 km s$^{-1}$, in good agreement with
the value of 140 km s$^{-1}$ for the Bar 1 region, where these objects are
located (Figure~\ref{veldistroregions}).

Almost all of the stars in our monitoring sample have $v_{\rm
  sys}$ within $2\sigma$ of the  mean.  However,
Star 5391 has $v_{\rm sys} = 44\ \rm km\ s^{-1}$, which is blueshifted
by 87 km s$^{-1}$, more than 3$\sigma$ from the mean and thus potentially a runaway star.  
This O8.5~III
star is also identified as a binary by our three methods
(Table~\ref{t_binaries}).

\subsection{Emission-line stars}
\label{estars}

A large fraction of our RIOTS4 stars turn out to be emission-line
stars, mostly classical Oe/Be stars.  We also identify four B supergiant
stars that exhibit forbidden emission lines \citep{Graus12}.  One of
these, star 29267 (AzV 154; \citealt{Azzopardi75})  was a
previously known sgB[e] star \citep{Zickgraf89}.  The other three
stars are 46398, 62661, and 83480 (R15, R38, and R48, respectively; \citealt{Feast60}).
SgB[e] stars are normally defined as stars exhibiting forbidden emission
lines along with strong IR dust excess.  However, this strong dust emission
is not present in the three RIOTS4 stars newly shown to be B[e] stars.
In \citet{Graus12}, we discuss these objects in
detail, demonstrating that they do show more modest, free-free near-IR
emission.  We propose that they represent a
new, transition class of dust-poor sgB[e] stars. 

\begin{figure*}
	\begin{center}
	\includegraphics[scale=1]{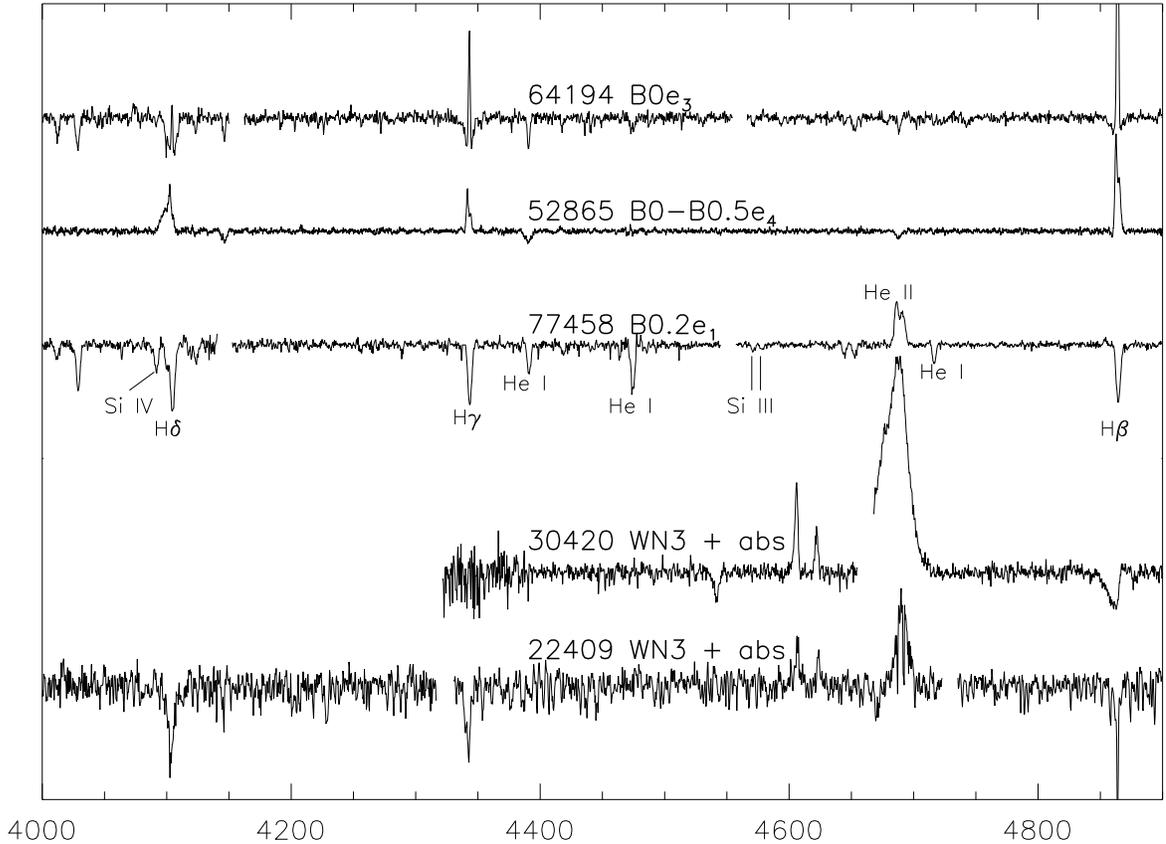}
	\caption{Spectra of the Wolf-Rayet and Be/X-ray binary stars in the RIOTS4 catalog.
	 }
	\label{WR-HMXB}
	\end{center}
\end{figure*}

There are two Wolf-Rayet stars included in the RIOTS4 survey.  They are
stars 22409 and 30420, which are both identified as WN3 + abs stars
by \citet{Massey01}.  In our RIOTS4 spectra, we detect only H
absorption lines for 22409, while 30420 also exhibits He
{\footnotesize II} absorption (Figure \ref{WR-HMXB}).  \citet{Massey01} identify He
{\footnotesize II} absorption in both objects and use the lack of He
{\footnotesize I} to estimate that the absorption components
correspond to O3-O4 stars.

The rest of the emission-line stars are classical Oe/Be stars, comprising
$\sim 25\%$ of the O stars \citep{GoldenMarx15} and $\sim 50\%$
of the B stars in the RIOTS4 survey.  These objects 
exhibit emission in their Balmer lines due to `decretion
disks' of material that are likely caused by rapid stellar rotation
\citep[e.g.,][]{Porter03}.  Oe/Be stars are more
common at lower metallicities, with a Galactic Oe/O-star fraction
of $0.03 \pm 0.01$ as measured from Galactic O Star Spectroscopic
Survey \citep[GOSSS;][]{Sota11, Sota14} and a $0.24 \pm 0.09$ Oe/O-star fraction
in SMC clusters \citep{Martayan10}.  The denominators here represent
all O stars, including Oe stars.  Similarly, the Be/B frequency of
$30-40\%$ in SMC clusters is about twice the Galactic frequency
\citep{Maeder99,Wisniewski06}.  This metallicity effect is consistent
with the decretion disk scenario, since the metal-poor SMC stars have
weak stellar winds, thereby impeding the loss of stellar angular
momentum through the winds.  The high rotation rates therefore
promote the formation of decretion disks, leading to the Be
phenomenon.

Our RIOTS4 field Oe stars and their
statistics are presented by \citet{GoldenMarx15}, yielding an Oe/O
ratio of $0.27\pm0.04$.  We also find that the Oe spectral
type distribution extends to earlier types than in the Galaxy, both in
terms of conventional classifications and hot Oe stars whose
spectral types are uncertain but are apparently of extremely
early type.  One extreme star, 77616, has He {\sc ii} in
emission from the disk, showing that even the hottest O stars can
present the Oe/Be phenomenon; this supports theoretical models
predicting that fast rotators can reach higher effective temperatures \citep{Brott11}.
Our large sample of Oe stars in the SMC strongly supports
the metallicity effects predicted by the decretion disk model and
characterizes the properties of early Oe stars.

Regarding the Be stars, the RIOTS4 Be/B fraction
appears to be even higher than found in previous studies.  This result
should be treated with caution because our sample selection criteria
may be biased to favor selection of Be stars.  These objects emit
strongly in H$\alpha$, which results in a brightening of their $R$
magnitude, thus lowering $Q_{UBR}$.  Therefore, our sample selection
criterion of $Q_{UBR} \leq -0.84$ is especially useful for selecting
Be stars.  Given our additional limiting $B$ magnitude criterion, it is unclear
whether our completeness limit for Be  stars extends to later spectral
types than normal stars, or whether it provides more complete
identification of B stars by including more Be stars down to the
magnitude limit.  A comprehensive treatment of the Be stars, 
including detailed investigation of the selection effects and
estimates of the luminosity classes, will be
presented in a future publication.  For now, we include \citet{Lesh68} classifications 
(Table \ref{catalog}) for these stars, which are a measure of the magnitude 
 of the Be phenomenon, and also indicate the presence of Fe II emission.   In total, the Oe/Be stars
account for 157 of the 374 stars (42\%) in the RIOTS4 sample.

We also observed three previously known Be/X-ray binary systems within
our survey, whose spectra are plotted in Figure \ref{WR-HMXB}.  Object 52865 is reported to be a
B0-0.5 III-Ve star in a binary system with a $967$-s pulsar
\citep{Schurch07, Haberl08}, and our spectral type for 52865 agrees
with this spectral classification.  \citet{Coe12} report object 64194
to be a B0.5-1 Ve star in a binary system with a presumed neutron
star, although no pulsar has been identified; we find a
spectral type of B0e$_3$ for this star.  Object 77458 is an eclipsing
X-ray binary with a period of $\sim3.9$ days \citep{Schreier72}.
\citet{Webster72} first identified 77458 as the optical counterpart of
the X-ray source, a $0.72$-s pulsar \citep{Lucke76} with a period of
3.9 days.  There is a variety of spectral classifications for 77458 in the literature, ranging
from O9.5 II \citep{Lennon97} to B2 I \citep{Garmany87}, which suggest
some real variation in this object's spectrum.  The most
recently published spectral type is O9.7 Ia+ from \citet{Evans04};
our spectral type for this object is slightly later, B0.2e$_1$. 

\section{Discussion}

The RIOTS4 survey provides a first, quantitative characterization of
the field massive star population based on a complete, uniformly
selected sample of OB stars.  It is also the first complete
survey of field massive stars in an external galaxy.  The
resulting characterization of this population is
necessarily sensitive to our definition of field stars,
recalling that our criterion requires that members be
at least 28 pc from other OB candidates, regardless of the presence of
lower-mass stars.  Thus, most of our objects can be expected to
represent the ``tip of the iceberg'' on low-mass clusters.  On the
other hand, we note that our 28-pc requirement is a more
stringent criterion for isolation than is often used in other
studies.  This clustering length is derived from the
spatial distribution of the entire OB population and represents a
characteristic value for the SMC as a galaxy \citep{Oey04}.
In contrast, other studies often use more arbitrary definitions,
for example, ``field'' OB stars in the vicinity of the 30 Doradus
giant star cluster \citep{Bressert12} correspond to a different
concept of field stars. 

\citet{Oey04} showed that the clustering law for SMC OB stars follows
an $N_*^{-2}$ power law extending down to $N_*=1$, which
corresponds to our individual RIOTS4 field OB stars, where $N_*$ is the
number of OB stars per cluster.  This basically
confirms that most of our sample corresponds to the
``tip of the iceberg'' objects, as expected.  However, as discussed in
detail by \citet{Oey04}, the magnitude of the $N_*=1$ bin does suggest
a slight, but difficult to quantify, enhancement above a simple
extrapolation of the power law distribution.  Conservatively, it is $<
0.3$ dex, implying that any excess ``deep field'' population is less
than 50\% of the total, and perhaps much less.

Based on Monte Carlo simulations, \citet{Lamb10} find that
observations of sparse clusters and field O stars are consistent with
full sampling of the IMF up to the universal stellar upper-mass limit
$m_{\rm up}$.  This is at odds with the steep upper IMF for the field stars
found in \S 4.2.  However, these results can be reconciled by the
fact that \citet{Lamb10} also identified the existence of an effective lower limit
to $N_*$ for normal clusters, $N_*\gtrsim 40$.  This value
corresponds to a mean cluster mass limit of $M_{\rm cl} \gtrsim 20\ M_\odot$
for a \citet{Kroupa01} IMF.  Since typically $m_{\rm up} \gtrsim M_{\rm cl}$ here, it
is apparent that in this regime it becomes physically impossible,
{\it on average,} to fully sample the IMF up to $m_{\rm up}$.  Therefore, the maximum
stellar masses in the sparsest clusters must necessarily be lower, on
average, than in normal clusters.  {\it Since our RIOTS4 sample is dominated by such
stars in sparse clusters, the steeper IMF is a natural consequence.}
This effect also provides a natural explanation for the value of the
steeper Salpeter IMF slope in clusters, compared with a simple --2
power law expected from simple Bondi-Hoyle accretion \citep{Oey11}.

Our RIOTS4 field stars therefore consist of both ``tip of the
iceberg'' stars that dominate small, but normal, clusters and ``deep
field'' objects that are substantially more isolated.  The former
correspond to objects that are consistent with stochastic sampling of
the IMF and clustering mass function, as described above; while the
latter correspond to objects that formed in greater isolation, if any,
and runaway stars. 

As discussed in \S 4.6, \citet{Mason09} estimate, based on somewhat
uncertain statistics, that Galactic field stars have a binary
frequency of about 51\%, as compared with a cluster frequency of
75\%.  If we assume that binaries actually form with the same
frequency in the field and clusters (i.e., 75\%), then the lower
observed field frequency can be attributed entirely to dilution by runaway stars,
which increase the number of single stars.  While dynamical ejection
mechanisms do predict some binary runaways, these should be relatively
insignificant for our purposes.  These assumptions imply that runaways
comprise 1/3 of all the massive field stars.  If we further assume
to first order that the tip-of-the-iceberg stars comprise another 50\%
of the field, as described above, then the remaining 1/6 of the sample
corresponds to objects that formed in extreme isolation.
We stress that these estimates are subject to substantial,
unknown uncertainties, and they only represent a first attempt at
understanding the field partition between these components.  For
example, if the runaway frequency is less than 33\%, this implies that
field stars actually form with a lower binary frequency in
the field.

Thus, the possibility remains that on the order of 1/6 of the field OB
stars may constitute a population that formed in extreme, or even
complete, isolation.   As described in \S 4.3, \citet{Oey13} presented
14 candidate field stars that appear to have formed in situ,
and 5 of these remain candidate members of this extremely isolated class.
\citet{Lamb10} also presented 3 such isolated candidate objects.  Thus we have at
least 8 candidate in-situ, deep-field OB stars, which may be around 13\% of
all such objects in our sample, based on the crude estimate above of
their contribution to the RIOTS4 survey.
As noted earlier, a number of other studies have also identified
strong candidates for isolated OB star formation in the Magellanic
Clouds and the Galaxy \citep[e.g.,][]{Selier11,Bressert12,Oskinova13}.

In \S~\ref{S_Runaways}, we found a lower limit to the runaway
frequency of $\sim 11$\%, which is consistent with our estimate of
$\sim 33$\%, following our analysis above.  Also, the
steep upper IMF (\S 4.2) suggests that runaway stars do not
dominate the field population.  Since O stars have a higher
observed runaway frequency than early B stars \citep{Gies87, Stone91},
the presence of runaways counteracts the IMF steepening discussed
earlier.  This is also consistent with the relatively high binary frequency
($0.59\pm 0.12$; \S 4.6.4) in our monitoring subsample. 

Thus, the picture of the field stellar population that emerges from
RIOTS4 and its ancillary studies is one that is dominated by ``tip of
the iceberg'' clusters, but with a significant fraction, on the order
of one-third, of runaway stars.  There is also evidence consistent with a
significant contribution, perhaps $\sim$17\%, from stars formed in extreme isolation.
Work is currently in progress to evaluate the relative contributions
of these components in the RIOTS4 survey.  At present, the
evidence remains consistent with highly isolated OB star formation
constituting a small fraction of the deep field.

\section{Conclusions}

The Runaways and Isolated O-Type Star Spectroscopic Survey of the SMC
(RIOTS4) provides a spatially complete, spectroscopic dataset for 
the field massive stars in the Small Magellanic Cloud obtained from
uniform criteria applied to the entire star-forming body of this
galaxy.  This survey sample is identified using photometric selection
criteria combined with a friends-of-friends algorithm to
identify the most isolated objects  \citep{Oey04}.  Over the 
course of five
years, we obtained spectra for all targets using
the IMACS and MIKE spectrographs on the Magellan
Telescopes.  From these spectra, we derive each star's spectral
classification and radial velocity.  

Using RIOTS4, we derived physical parameters such as
the stellar effective temperatures and masses, allowing us to
investigate the shape of the field IMF above $20M_\odot$
\citep{Lamb13}.  We find that the slope of the field
massive star IMF is significantly steeper ($\Gamma$=2.3$\pm 0.4$) than the
traditional Salpeter slope ($\Gamma$=1.35).  This result is consistent
with the $\Gamma$=1.8 IMF slope found by \citet{Parker98} and
qualitatively corroborates previous observations of a steep field IMF
slope of $\Gamma \sim 3-4$ in the Magellanic Clouds
\citep{Massey95,Massey02}.  Complete details are given by \citet{Lamb13}.   

We also use RIOTS4 data to probe limits of the most
massive stars that can form in isolation or within sparsely populated
clusters \citep{Lamb10,Oey13}.  In conjunction with {\sl HST} and
ground-based imaging, we identify sparse clusters associated with
target OB  stars in the RIOTS4
sample.  With cluster mass estimates and RIOTS4 stellar masses, we
examine the relationship between the most massive star in a cluster
and the mass of the parent cluster.  Our results are consistent
with cluster mass being independent of the most massive
member star.  This applies unless the total cluster masses are so
small that stars near the upper-mass limit cannot form.  This
suppression of the most massive stars in the smallest clusters
explains the steep field IMF observed above.
We also identify a compelling sample of candidate field
OB stars that may have formed in situ, given their apparent lack of
runaway velocities and central location within dense {\sc H ii} regions
\citep{Oey13}. 

We use the radial velocities of RIOTS4 stars to examine the 
large-scale velocity structure of the SMC, and for an initial look at 
the kinematics of the field OB population and runaway frequency.
We find that the kinematics mirror those of other
surveys of massive stars \citep{Evans08} and gas \citep{Stanimirovic04}.  We
find the systemic velocity of the SMC is $\sim 150$ km s$^{-1}$, with
a large velocity gradient as a function of position that roughly
follows the gradient observed in {\sc H i} gas \citep{Stanimirovic04}.   
Given this large velocity gradient, we must consider the
line-of-sight SMC systemic velocity as given by the gas kinematics
when identifying runaway stars within our survey.  Thus, we compare
the stellar radial velocity for each RIOTS4 star with the local {\sc H
  i} gas velocity in the line of sight from \citet{Stanimirovic99}.
Runaway candidates are defined to be
those objects with a difference $> 30$ km s$^{-1}$ between stellar and
{\sc H i} radial velocities.  We find that 11\% of the sample meets
this criterion, which is a lower bound due to our inability to detect transverse runaways.
The identification of a binary runaway system and a candidate high-velocity
(200 km s$^{-1}$) runaway star suggest that dynamical ejection is a
significant and possibly dominant contributor to the runaway OB population.

To identify binary stars within our sample, we look for stellar radial
velocity variations using $9-16$ epochs of data for three IMACS
multi-slit fields encompassing 29 stars.  We use three methods to identify binary stars.
First, binaries are likely to be those objects that exhibit large radial
velocity variations whose amplitudes correlate with time
interval.  Second, we identify binary candidates using a statistical
F-test, comparing the observed velocity variation with that expected from
observational uncertainties \citep{Duquennoy91}.  Third, we identify
candidates using the periodicity power spectrum and then fitting for
orbital solutions.  All three methods find 10 out of 17 normal OB stars
($59\% \pm 12\%$) to be strong binary candidates.  This can be
compared with the binary fraction found in Galactic clusters and OB
associations, which is $\sim 60-80\%$, and that for
Galactic field stars, which is $\sim 40-50\%$.

The RIOTS4 sample also includes a large number of emission-line stars,
including two Wolf-Rayet stars and a newly identified population of
dust-poor B[e] supergiant stars that may represent a transition class
of objects \citep{Graus12}.  The remainder of the emission-line stars
are classical Oe/Be stars, which occur at a higher frequency in the
SMC than in the Galaxy.  The RIOTS4 data clearly extend this
finding to early Oe stars and to field Oe/Be stars.  Our Oe/O-star frequency
of $0.27\pm0.04$ in the SMC is significantly greater than the Milky Way value, and
the SMC spectral type distribution also extends to the hottest
effective temperatures, in contrast to Milky Way objects \citep{GoldenMarx15}.  These results
support the decretion disk model for the Be phenomenon, since
metal-poor stars rotate faster due to their inability to remove angular
momentum via stellar winds.  Similarly, our frequency of Be/B stars is
higher than Galactic values, but this result may be biased
by our photometric selection criteria.  We will examine the RIOTS4 Be
stars in a future work.

Work is also underway to evaluate the fraction of deep field objects
relative to ``tip of the iceberg'' stars, which will further clarify 
the statistics of OB star formation in the sparsest regime.  In
addition, we have initiated follow-up spectroscopic monitoring to obtain binary
star properties, including systemic velocities.  These observations will yield
reliable statistics for runaway stars, data on $v\sin i$, and Oe/Be star variability.

\acknowledgments

Many individuals helped make this publication a reality, including the
referee, who provided thoughtful comments.
Thanks to Nidia Morrell and Phil Massey for
advice on radial velocity measurements, and to Thomas Bensby, Tom
Brink, and Jess Werk for advice on the data reduction pipelines.
Thanks to Mario Mateo for help with scheduling the binary monitoring
runs and observing advice.  We thank Fred Adams, Rupali Chandar, Xinyi
Chen, Oleg Gnedin, Lee Hartmann, Wen-hsin Hsu, Anne Jaskot, Mario
Mateo, Eric Pellegrini, and Jordan Zastrow for helpful discussions.   
This work was supported by the National Science Foundation grants
AST-0907758, AST-1514838; NASA grant NAG4-9248; and the University of Michigan,
Rackham Graduate School. 
\bibliography{./ms}

\end{document}


\begin{deluxetable}{cccccccccc}

 \tabletypesize{\small}
  \tablewidth{0pc}
  \tablecaption{RIOTS4 Catalog\tablenotemark{a}} 
  \tablehead{ \colhead{ID\tablenotemark{b}}
    & \colhead{$B$\tablenotemark{b}} & \colhead{$V$\tablenotemark{b}} &
  \colhead{$Q_{UBR}$} & \colhead{$A_V$\tablenotemark{c}} & \colhead{Sp Type} &
  \colhead{RV$_{\rm star}$ }&
  \colhead{RV$_{\rm HI}$\tablenotemark{d}}& \colhead{Instrument} &
  \colhead{Observation Date} \\
  &  &  &  &  &  & \colhead{(km s$^{-1}$)} & \colhead{(km s$^{-1}$)}
& & \colhead{  (YYMMDD)}
}
\startdata

107 & 14.96 & 15.00 & -0.95 & 0.82 &    Be$_3$        & -- & -- &     MIKE  &  111024 \\
298 & 15.18 & 15.12 & -0.91 & 1.03 &    B1e$_{3+}$    & -- & -- &     IMACS f/4  &  070920 \\
1037 & 15.15 & 15.28 & -0.85 & 0.44 &    B0.5 V   & 110 & 110 &     IMACS f/4  &  070920 \\
1600 & 14.42 & 14.60 & -0.87 & 0.32 &    O8.5 V   & 93 & 103 &   IMACS f/4  &  070920 \\
1631 & 15.19 & 15.15 & -0.99 & 1.11 &    B1e$_2$    & 120 & 120 &   IMACS f/4  &  070920 \\
1830 & 13.84 & 13.94 & -0.85 & 0.49 &    B0.5 III         & -- & -- &     IMACS f/4  &  070920 \\
1952 & 14.98 & 14.93 & -0.91 & 1.09 &    B1e$_2$    & 80 & 97 &   IMACS f/4  &  070920 \\
2034 & 14.51 & 14.57 & -1.01 & 0.78 &    B       & -- & -- &     MIKE  &  111024 \\
2093 & 15.12 & 15.19 & -0.95 & 0.68 &    B1e$_{3+}$    & -- & -- &     IMACS f/4  &  070920 \\
2666 & 15.15 & 15.18 & -0.88 & 0.70 &    B1.5e$_{3+}$          & 117 & 117 &   IMACS f/4  &  081007 \\
3173 & 14.66 & 14.37 & -1.73 & 0.99 &    O3-4 + neb        & 111 & 111 &   IMACS f/4  &  081011 \\
3224 & 14.86 & 14.70 & -0.92 & 1.39 &    B1e$_{2+}$    & 102 & 102 &   IMACS f/4  &  070920 \\
3459 & 13.32 & 13.46 & -0.93 & 0.48 &    O9.5 I   & 201 & 169 &   IMACS f/4  &  081011 \\
3722 & 15.20 & 14.83 & -1.08 & 0.46 &    B       & -- & -- &     IMACS f/4  &  081007 \\
3815 & 14.97 & 14.92 & -1.02 & 1.13 &    Be$_2$       & -- & -- &     IMACS f/4  &  081007 \\
4294 & 14.96 & 15.00 & -0.93 & 0.78 &    Be$_2$       & -- & -- &     MIKE  &  111024 \\
4424 & 14.70 & 14.82 & -0.90 & 0.49 &    B0 III           & -- & -- &     IMACS f/4  &  081011 \\
4919 & 13.66 & 13.85 & -0.95 & 0.33 &    B0 III   & 118 & 118 &     MIKE  &  111024 \\
5041 & 15.09 & 15.09 & -0.96 & 0.96 &    B0 V     & 130 & 130 &     MIKE  &  111024 \\
5063 & 14.66 & 14.60 & -0.87 & 1.02 &    B1e$_3$          & 116 & 116 &   IMACS f/4  &  081007 \\
5313 & 14.89 & 15.11 & -0.87 & 0.23 &    O8.5 V   & 135 & 135 &   IMACS f/4  &  081007 \\
5905 & 14.87 & 14.90 & -0.91 & 0.75 &    B0e$_2$    & -- & -- &     IMACS f/4  &  081007 \\
6908 & 14.77 & 14.53 & -1.02 & 1.71 &    O9.5-B0 III      & 128 & 128 &   IMACS f/4  &  081006\tablenotemark{e} \\
6940 & 14.25 & 14.33 & -0.96 & 0.68 &    B1e$_{3+}$    & 131 & 131 &   IMACS f/4  &  081011 \\
7254 & 14.75 & 14.74 & -0.86 & 0.86 &    O9.5 IIIe$_2$        & 126 & 126 &     IMACS f/4  &  081006\tablenotemark{e} \\
7329 & 14.32 & 14.45 & -0.94 & 0.49 &    Be$_2$       & -- & -- &     MIKE  &  111024 \\
7437 & 12.93 & 13.12 & -0.94 & 0.33 &    O6.5 I(f)        & 151 & 151 &   IMACS f/4  &  081006\tablenotemark{e} \\
7782 & 14.30 & 14.46 & -0.91 & 0.40 &    O8 V     & 127 & 127 &   IMACS f/4  &  081006\tablenotemark{e} \\
8098 & 14.13 & 14.17 & -1.00 & 0.85 &    B1e$_3$          & 165 & 165 &   MIKE  &  111024 \\
8609 & 13.98 & 14.11 & -0.87 & 0.48 &    B0 III           & 128 & 128 &   IMACS f/4  &  081006\tablenotemark{e} \\
9732 & 14.63 & 14.81 & -0.88 & 0.31 &    O7 Vz    & 136 & 136 &   IMACS f/4  &  081011 \\
10026 & 14.64 & 14.68 & -1.02 & 0.87 &    Be$_{3+}$       & -- & -- &     IMACS f/4  &  081011 \\
10421 & 14.81 & 14.99 & -0.87 & 0.28 &    B1 V     & 152 & 161 &   IMACS f/4  &  070919 \\
10556 & 14.60 & 14.63 & -0.89 & 0.74 &    B0.2 V   & 100 & 120 &   IMACS f/4  &  090825 \\
10671 & 14.88 & 14.89 & -0.84 & 0.85 &    B0.5 V   & 122 & 122 &   IMACS f/4  &  081006\tablenotemark{e} \\
11045 & 14.80 & 15.01 & -0.87 & 0.21 &    O8.5 V   & 182 & 169 &   IMACS f/4  &  070919 \\
11087 & 14.67 & 14.71 & -0.93 & 0.80 &    B1e$_1$    & -- & -- &     IMACS f/4  &  090825 \\
11213 & 14.88 & 14.77 & -0.97 & 1.29 &    Be$_{3+}$       & -- & -- &     IMACS f/4  &  090824 \\
11238 & 14.41 & 14.30 & -0.96 & 1.21 &    O + B      & -- & -- &     MIKE  &  111024 \\
11280 & 14.15 & 14.16 & -0.94 & 0.90 &    B1e$_3$     & 125 & 138 &   IMACS f/4  &  070919 \\
11623 & 14.12 & 14.13 & -0.86 & 0.83 &    O9 III:          & 170 & 170 &   IMACS f/4  &  060912 \\
11677 & 14.47 & 14.46 & -1.01 & 1.04 &    O9 III:e$_{3+}$         & -- & -- &     IMACS f/4  &  081011 \\
11777 & 13.76 & 13.88 & -0.92 & 0.53 &    B0.2 V   & 168 & 168 &   IMACS f/4  &  060912 \\
11802 & 14.62 & 14.59 & -0.86 & 0.96 &    B0.2 IV          & 155 & 155 &   IMACS f/4  &  090824 \\
12102 & 14.64 & 14.69 & -0.98 & 0.84 &    O9 IIIe$_2$         & -- & -- &     IMACS f/4  &  081011 \\
13075 & 12.93 & 13.05 & -0.89 & 0.52 &    O9.5 I   & 131 & 131 &   IMACS f/4  &  060912 \\
13314 & 14.76 & 14.77 & -0.87 & 0.84 &    B1 III   & 168 & 215 &     IMACS f/4  &  070919 \\
13682 & 14.94 & 14.92 & -0.96 & 1.03 &    Be$_3$        & -- & -- &     IMACS f/4  &  090825 \\
13774 & 15.07 & 15.07 & -0.84 & 0.85 &    --       & -- & -- &     IMACS f/4  &  060913 \\
13831 & 13.91 & 14.06 & -0.87 & 0.39 &    B0.5 V   & 172 & 172 &   IMACS f/4  &  060912 \\
13896 & 13.56 & 13.76 & -0.98 & 0.30 &    O8 III((f))      & 159 & 163 &   IMACS f/4  &  081011 \\
14324 & 14.17 & 14.11 & -0.97 & 1.12 &    O6 V((f))e$_2$       & 152 & 152 &   IMACS f/4  &  090824 \\
14878 & 14.11 & 14.25 & -0.89 & 0.46 &    O9 III   & 147 & 147 &   IMACS f/4  &  090825 \\
15060 & 14.11 & 14.29 & -0.88 & 0.30 &    B0.2 III-IV     & 155 & 155 &   IMACS f/4  &  060912 \\
15102 & 15.17 & 15.01 & -0.89 & 1.40 &    B1.5 V   & 103 & 123 &   IMACS f/4  &  081011 \\
15256 & 15.09 & 15.00 & -0.90 & 1.14 &    B1 V     & 157 & 157 &   IMACS f/4  &  060912 \\
15263 & 15.05 & 15.22 & -0.87 & 0.28 &    B0.2 V   & 109 & 156 &   IMACS f/4  &  070919 \\
15271 & 13.36 & 13.54 & -0.94 & 0.39 &    O6 III((f))e$_1$     & 143 & 153 &   IMACS f/4  &  070919 \\
15618 & 14.84 & 14.95 & -0.92 & 0.53 &    B        & 128 & 128 &   IMACS f/4  &  090825 \\\
15742 & 13.51 & 13.64 & -0.92 & 0.53 &    O9 III   & 168 & 168 &   IMACS f/4  &  090825 \\
15854 & 15.21 & 14.51 & -1.17 & 0.68 &    B1e$_3$          & 107 & 117 &   IMACS f/4  &  090824 \\
16039 & 14.53 & 14.54 & -1.03 & 0.99 &    B1e$_{4+}$    & -- & -- &     IMACS f/4  &  090825 \\
16147 & 14.03 & 14.15 & -0.91 & 0.53 &    B0 V     & 147 & 147 &   IMACS f/4  &  060912 \\
16230 & 13.01 & 13.20 & -0.90 & 0.29 &    O9 III:          & 71 & 136 &   IMACS f/4  &  060912 \\
16481 & 14.16 & 14.26 & -0.94 & 0.62 &    O9.5 V   & 160 & 160 &   IMACS f/4  &  070919 \\
16518 & 13.77 & 13.90 & -0.96 & 0.53 &    B0 V     & 176 & 171 &   IMACS f/4  &  070919 \\
16587 & 13.99 & 13.91 & -1.00 & 1.24 &    B0 V     & -- & -- &     IMACS f/4  &  070920 \\
16616 & 14.18 & 14.29 & -0.86 & 0.50 &    --       & -- & -- &     IMACS f/4  &  060913 \\
17240 & 14.37 & 14.47 & -0.85 & 0.55 &    O7.5 V   & 105 & 108 &   IMACS f/4  &  060912 \\
17813 & 14.83 & 14.84 & -0.86 & 0.86 &    B0 V     & 192 & 184 &   IMACS f/4  &  070920 \\
18187 & 14.95 & 15.00 & -0.95 & 0.79 &    Be$_{2+}$        & -- & -- &     IMACS f/4  &  070919 \\
18301 & 14.40 & 14.58 & -0.86 & 0.31 &    B0.2 V   & 155 & 155 &   IMACS f/4  &  070919 \\
18329 & 14.64 & 14.65 & -0.98 & 0.91 &    O9.5 IIIe$_{4+}$ pec   & 119 & 123 &   IMACS f/4  &  070920 \\
18373 & 13.48 & 13.41 & -1.05 & 1.26 &    Be$_3$        & 160 & 160 &   MIKE  &  111025 \\
19382 & 14.27 & 14.31 & -0.96 & 0.81 &    B0 V     & 153 & 153 &   IMACS f/4  &  070920 \\
19728 & 12.49 & 12.59 & -0.89 & 0.59 &    B1 I          & 166 & 174 &   IMACS f/4  &  070919 \\
20939 & 14.14 & 14.18 & -0.86 & 0.73 &    B1 II    & 155 & 155 &   IMACS f/4  &  060912 \\
21844 & 14.09 & 14.18 & -0.86 & 0.57 &    O8 III((f))      & 151 & 151 &   IMACS f/4  &  080920\tablenotemark{e} \\
21877 & 14.96 & 15.10 & -0.95 & 0.46 &    O7 V     & 91 & 113 &   IMACS f/4  &  060912 \\
21933 & 15.08 & 14.83 & -0.92 & 1.69 &    Be$_3$       & 130 & 130 &     IMACS f/4  &  081006\tablenotemark{e} \\
21983 & 15.18 & 15.18 & -0.98 & 0.96 &    O9.5 V   & 174 & 174 &   IMACS f/4  &  070920 \\
22178 & 13.34 & 13.49 & -0.91 & 0.43 &    O + B      & 176 & 171 &   MIKE  &  111025 \\
22321 & 13.70 & 13.69 & -0.98 & 1.00 &    O9.5 IIIpe$_{4+}$ & 167 & 167 &     IMACS f/4  &  080920\tablenotemark{e} \\
22409 & 15.14 & 14.97 & -1.01 & 1.52 &    WN3 + abs\tablenotemark{f}         & 98 & 118 &   IMACS f/4  &  060912\\
22451 & 13.98 & 14.14 & -0.93 & 0.41 &    O9 III   & -- & -- &     IMACS f/4  &  070920 \\
22623 & 15.19 & 15.21 & -0.86 & 0.84 &    B0e$_3$          & -- & -- &     IMACS f/4  &  070920 \\
23710 & 14.85 & 14.74 & -0.97 & 1.31 &    O9-B0 pe$_{3+}$ & 168 & 168 &     IMACS f/4  &  080920\tablenotemark{e} \\
23859 & 14.75 & 14.82 & -0.93 & 0.72 &    B1e$_3$    & -- & -- &     IMACS f/4  &  060912 \\
23954 & 14.59 & 14.58 & -0.85 & 0.86 &    B1.5e$_{3+}$          & 130 & 130 &   IMACS f/4  &  080920\tablenotemark{e} \\
24056 & 13.61 & 13.64 & -0.85 & 0.76 &    B1.5 III         & 199 & 186 &   IMACS f/4  &  070920 \\
24096 & 14.56 & 14.53 & -1.05 & 1.07 &    B0 V     & 138 & 135 &   IMACS f/4  &  060912 \\
24119 & 14.78 & 14.80 & -1.01 & 0.88 &    B0-0.2 V         & -- & -- &     IMACS f/4  &  070920 \\
24213 & 14.17 & 14.34 & -0.88 & 0.31 &    B0 III           & 126 & 126 &   IMACS f/4  &  080920\tablenotemark{e} \\
24229 & 14.79 & 14.80 & -0.86 & 0.82 &    B1e$_2$    & 155 & 155 &     IMACS f/4  &  080920\tablenotemark{e} \\
24895 & 14.23 & 14.20 & -0.88 & 0.99 &    Be$_3$       & -- & -- &     MIKE  &  111024 \\
24914 & 14.19 & 14.19 & -1.00 & 0.95 &    O9 III-Vpe$_1$       & 81 & 105 &     IMACS f/4  &  081121\tablenotemark{e} \\
25282 & 14.73 & 14.74 & -0.90 & 0.90 &    B0e$_1$    & 130 & 130 &   IMACS f/4  &  080920\tablenotemark{e} \\
25337 & 15.12 & 15.03 & -0.89 & 1.12 &    Be$_3$       & 124 & 124 &     IMACS f/4  &  080920\tablenotemark{e} \\
25639 & 14.85 & 14.69 & -0.96 & 1.42 &    B0.5e$_{3+}$          & -- & -- &     IMACS f/4  &  060912 \\
25974 & 15.08 & 15.08 & -0.85 & 0.75 &    B1.5 V   & 116 & 123 &   IMACS f/4  &  060912 \\
27135 & 15.20 & 15.32 & -0.93 & 0.56 &    B1e$_2$    & 113 & 117 &   IMACS f/4  &  080920\tablenotemark{e} \\
27496 & 14.60 & 14.77 & -0.84 & 0.27 &    B0.5 V   & 137 & 137 &   IMACS f/4  &  060912 \\
27600 & 13.79 & 13.90 & -0.89 & 0.53 &    B0.5 III         & 177 & 177 &   IMACS f/4  &  080920\tablenotemark{e} \\
27712 & 14.96 & 15.09 & -0.85 & 0.44 &    B1.5 V   & 127 & 127 &   IMACS f/4  &  080920\tablenotemark{e} \\
27736 & 14.54 & 14.54 & -0.99 & 0.95 &    B0e$_2$    & 153 & 153 &   IMACS f/4  &  080920\tablenotemark{e} \\
27884 & 14.35 & 14.35 & -1.06 & 1.05 &    O7-8.5 Vpe$_{4+}$    & 156 & 156 &     IMACS f/4  &  080920\tablenotemark{e} \\
28496 & 15.01 & 15.07 & -0.87 & 0.73 &    B0 V     & 154 & 154 &   MIKE  &  111024 \\
28841 & 14.24 & 14.27 & -0.89 & 0.78 &    B1 III   & 141 & 141 &   IMACS f/4  &  080920\tablenotemark{e} \\
29267 & 13.48 & 13.20 & -1.01 & 1.78 &    B[e]     & -- & -- &     MIKE  &  111024 \\
29312 & 15.01 & 15.29 & -0.87 & 0.00 &    B0.7 III         & 162 & 162 &   IMACS f/4  &  060912 \\
30018 & 14.83 & 14.95 & -0.93 & 0.48 &    B0.2 III         & 160 & 160 &   MIKE  &  111024 \\
30420 & 15.12 & 15.24 & -0.97 & 0.56 &    WN3 + abs\tablenotemark{f}     & -- & -- &     IMACS f/4  &  060912 \\
30472 & 14.55 & 14.67 & -0.87 & 0.48 &    B1e$_1$           & -- & -- &     IMACS f/4  &  060912 \\
30492 & 14.45 & 14.51 & -0.92 & 0.72 &    B0.5          & 179 & 179 &   MIKE  &  111024 \\
30744 & 14.24 & 14.30 & -0.88 & 0.66 &    B1 III + O9.5 Ve$_2$         & -- & -- &     IMACS f/4  &  060912 \\
31574 & 15.17 & 15.10 & -0.90 & 1.05 &    B1 V     & 149 & 145 &   IMACS f/4  &  060912 \\
31699 & 14.73 & 14.78 & -0.90 & 0.73 &    B1e$_1$        & -- & -- &     IMACS f/4  &  060912 \\
32449 & 14.19 & 14.29 & -0.86 & 0.51 &    B1.5 III         & 121 & 121 &   IMACS f/4  &  060912 \\
32552 & 14.85 & 14.87 & -0.96 & 0.88 &    B0e$_3$    & -- & -- &     IMACS f/4  &  090825 \\
32752 & 14.63 & 14.57 & -0.90 & 1.04 &    Be$_{3+}$        & 270 & 186 &   MIKE  &  111025 \\
33135 & 15.20 & 15.26 & -0.95 & 0.77 &    B0e$_3+$     & -- & -- &     IMACS f/4  &  090825 \\
33245 & 15.10 & 15.23 & -0.91 & 0.47 &    Be$_1$        & -- & -- &     IMACS f/4  &  060912 \\
33823 & 14.45 & 14.62 & -0.89 & 0.35 &    O9.5 III         & 170 & 170 &   MIKE  &  111025 \\
34005 & 14.80 & 15.01 & -1.02 & 0.29 &    O9.5 V   & 166 & 166 &   IMACS f/4  &  090825 \\
34315 & 13.68 & 13.78 & -0.88 & 0.54 &    B1 II    & 138 & 138 &   IMACS f/4  &  060912 \\
34457 & 13.74 & 13.93 & -0.91 & 0.30 &    B0 III   & 168 & 168 &   MIKE  &  111025 \\
34651 & 15.01 & 14.99 & -0.87 & 0.92 &    Be$_1$         & -- & -- &     MIKE  &  111025 \\
34988 & 14.84 & 14.85 & -0.95 & 0.89 &    Be$_2$       & 226 & 183 &   MIKE  &  111025 \\
35474 & 14.56 & 14.45 & -0.93 & 1.53 &    B1 III   & 224 & 184 &   MIKE  &  111025 \\
35491 & 14.66 & 14.71 & -0.90 & 0.77 &    O8 V     & 126 & 126 &   IMACS f/4  &  060911 \\
35598 & 14.88 & 15.15 & -0.94 & 0.05 &    O8 V     & 162 & 162 &   IMACS f/4  &  060911 \\
36175 & 12.64 & 12.68 & -0.85 & 0.73 &    B1 II    & 191 & 191 &   MIKE  &  111024 \\
36213 & 13.21 & 13.38 & -0.90 & 0.38 &    O9.5 III + B        & 120 & 120 &   MIKE  &  111025 \\
36294 & 14.51 & 14.55 & -0.95 & 0.80 &    B0.5e$_1$           & 130 & 130 &   IMACS f/4  &  060911 \\
36325 & 15.02 & 14.88 & -0.88 & 1.33 &    O9.5 V   & 146 & 146 &   IMACS f/4  &  060912 \\
36514 & 15.01 & 15.15 & -0.87 & 0.47 &    O9 V     & 141 & 141 &   IMACS f/4  &  060911 \\
36815 & 14.93 & 14.98 & -0.96 & 0.78 &    Be$_2$        & 155 & 155 &   MIKE  &  111025 \\
36975 & 15.04 & 15.04 & -0.89 & 0.87 &    Be$_{4+}$        & 164 & 164 &   MIKE  &  111026 \\
37419 & 15.12 & 14.95 & -0.98 & 1.54 &    Be$_{4+}$         & 127 & 127 &   IMACS f/4  &  060911 \\
37502 & 14.65 & 14.62 & -1.03 & 1.10 &    O9.5-B0: pe$_{3+}$      & -- & -- &     IMACS f/4  &  060911 \\
38024 & 14.32 & 14.53 & -0.94 & 0.27 &    O4 V((f))        & 103 & 151 &   IMACS f/4  &  090825 \\
38036 & 14.28 & 14.33 & -0.98 & 0.82 &    O6.5-7: Vpe$_3$       & 152 & 152 &   IMACS f/4  &  060911 \\
38445 & 14.67 & 14.68 & -1.09 & 1.02 &    Be$_{4+}$         & -- & -- &     IMACS f/4  &  090825 \\
38508 & 15.14 & 14.95 & -0.96 & 1.52 &    B0.7          & 179 & 179 &   IMACS f/4  &  090825 \\
38893 & 14.74 & 14.89 & -0.90 & 0.41 &    B0e$_1$    & 179 & 179 &   IMACS f/4  &  060911 \\
38921 & 14.07 & 14.12 & -0.96 & 0.80 &    B0 III           & 154 & 154 &   IMACS f/4  &  060911 \\
39211 & 14.73 & 14.94 & -0.86 & 0.18 &    O9.5 III         & 156 & 156 &   IMACS f/4  &  090825 \\
39904 & 15.13 & 15.17 & -0.98 & 0.81 &    B0e$_2$    & -- & -- &     IMACS f/4  &  090825 \\
40380 & 14.22 & 14.33 & -0.92 & 0.58 &    O7 V     & 181 & 178 &   IMACS f/4  &  090825 \\
40504 & 13.98 & 14.15 & -0.87 & 0.30 &    B1.5 III         & 95 & 150 &   IMACS f/4  &  090825 \\
41183 & 14.78 & 14.84 & -0.95 & 0.77 &    Be$_1$        & -- & -- &     IMACS f/4  &  101222 \\
41345 & 14.87 & 15.04 & -0.89 & 0.34 &    B1 III       & -- & -- &     IMACS f/4  &  060911 \\
41648 & 13.86 & 13.95 & -0.90 & 0.61 &    B0.5 III         & 106 & 118 &   IMACS f/4  &  060911 \\
42260 & 14.43 & 14.54 & -0.85 & 0.50 &    O9.5 III         & 140 & 140 &   IMACS f/4  &  060911 \\
42654 & 14.84 & 14.78 & -0.99 & 1.17 &    Be$_{3+}$       & -- & -- &     IMACS f/4  &  090825 \\
42959 & 14.59 & 14.72 & -0.91 & 0.48 &    B        & -- & -- &     --  &      -- \\
43411 & 14.90 & 14.97 & -0.98 & 0.72 &    O9 V     & 132 & 150 &   IMACS f/4  &  081006 \\
43589 & 14.93 & 15.04 & -0.89 & 0.56 &    B0e$_3$        & -- & -- &     IMACS f/4  &  060911 \\
43724 & 14.05 & 14.25 & -0.96 & 0.32 &    O7 III((f))      & 100 & 148 &   IMACS f/4  &  101222 \\
44316 & 15.04 & 15.15 & -0.93 & 0.58 &    Be$_3$       & 192 & 169 &   MIKE  &  111026 \\
44336 & 14.78 & 14.72 & -1.01 & 1.20 &    B1.5e          & -- & -- &     IMACS f/4  &  090825 \\
44965 & 14.65 & 14.69 & -1.06 & 0.90 &    B0e$_2$          & -- & -- &     IMACS f/4  &  070920 \\
45640 & 13.95 & 14.01 & -0.97 & 0.74 &    B1e$_2$           & 142 & 150 &   IMACS f/4  &  081006 \\
46022 & 14.74 & 14.85 & -0.84 & 0.52 &    O9.5 V   & 148 & 153 &   IMACS f/4  &  101221 \\
46035 & 14.05 & 14.24 & -0.95 & 0.34 &    O7 V     & -- & -- &     IMACS f/4  &  101221 \\
46090 & 14.58 & 14.66 & -0.88 & 0.64 &    B0e$_2$    & -- & -- &     IMACS f/4  &  070920 \\
46241 & 15.17 & 15.11 & -0.86 & 1.00 &    B1 V     & 117 & 117 &   IMACS f/4  &  060912 \\
46317 & 14.53 & 14.76 & -0.90 & 0.17 &    O8.5 V   & 174 & 174 &   IMACS f/4  &  070920 \\
46392 & 14.39 & 14.56 & -0.93 & 0.35 &    B0.7 IV          & 138 & 156 &   IMACS f/4  &  101222 \\
46398 & 12.67 & 12.64 & -0.94 & 1.00 &    B[e]     & -- & -- &     MIKE  &  111024 \\
46831 & 15.05 & 15.15 & -0.87 & 0.54 &    O7.5-9 V         & 145 & 148 &   IMACS f/4  &  081006 \\
47029 & 15.08 & 15.18 & -0.84 & 0.53 &    B0.2 V   & 188 & 188 &   IMACS f/4  &  101221 \\
47459 & 15.10 & 15.17 & -0.85 & 0.55 &    B0 V  + B     & 155 & 155 &   IMACS f/4  &  060912 \\
47478 & 14.74 & 14.96 & -0.98 & 0.28 &    O7 V     & 143 & 154 &   IMACS f/4  &  101222 \\
47668 & 15.04 & 15.16 & -0.97 & 0.56 &    Be       & -- & -- &     --  &      -- \\
47908 & 14.64 & 14.19 & -1.10 & 0.52 &    Be       & -- & -- &     IMACS f/4  &  090824 \\
48037 & 14.56 & 14.54 & -1.00 & 1.03 &    B0e$_3$    & -- & -- &     IMACS f/4  &  090824 \\
48057 & 14.91 & 14.93 & -0.97 & 0.87 &    B0.2e$_{3+}$        & -- & -- &     IMACS f/4  &  090824 \\
48170 & 14.16 & 14.29 & -0.96 & 0.53 &    O9.5 V   & 145 & 145 &   IMACS f/4  &  081011 \\
48266 & 15.19 & 15.17 & -1.08 & 1.06 &    B0e$_{3+}$    & -- & -- &     IMACS f/4  &  081006 \\
48432 & 14.94 & 15.07 & -0.90 & 0.37 &    B0III + B          & 201 & 186 &   MIKE  &  111025 \\
48601 & 13.58 & 13.71 & -0.92 & 0.49 &    B0 III           & 133 & 133 &   IMACS f/4  &  060912 \\
48882 & 15.19 & 15.16 & -0.99 & 1.04 &    B1e$_{3+}$    & 148 & 158 &   IMACS f/4  &  090824 \\
49014 & 14.53 & 14.60 & -0.95 & 0.69 &    B0e$_{3+}$   & 133 & 133 &   IMACS f/4  &  090824 \\
49450 & 14.07 & 14.22 & -0.86 & 0.37 &    O9.5 V   & 150 & 156 &   IMACS f/4  &  101222 \\
49517 & 15.02 & 14.97 & -0.96 & 1.11 &    B1.5 V   & 138 & 138 &   IMACS f/4  &  060912 \\
49580 & 14.22 & 14.38 & -0.88 & 0.39 &    O9.5 IV          & 124 & 124 &   IMACS f/4  &  090824 \\
49825 & 14.75 & 14.94 & -0.88 & 0.29 &    B0.2 III         & 140 & 156 &   IMACS f/4  &  101222 \\
49937 & 15.14 & 15.19 & -0.90 & 0.74 &    B0V      & 395 & 192 &   MIKE  &  111025 \\
50095 & 14.48 & 14.51 & -0.94 & 0.87 &    O9 Ve$_{4+}$   & -- & -- &     IMACS f/4  &  060912 \\
50153 & 14.64 & 14.46 & -1.47 & 1.92 &    H II region       & -- & -- &     MIKE  &  111025 \\
50331 & 12.83 & 12.99 & -0.92 & 0.41 &    --       & -- & -- &     IMACS f/4  &  060912 \\
50396 & 15.17 & 15.16 & -1.00 & 1.00 &    Be$_2$       & -- & -- &     IMACS f/4  &  090824 \\
50609 & 12.43 & 12.55 & -0.90 & 0.53 &    B1 I     & 149 & 141 &   IMACS f/4  &  090824 \\
50791 & 15.04 & 15.19 & -0.93 & 0.47 &    O8 V     & 190 & 183 &   IMACS f/4  &  081007 \\
50825 & 14.58 & 14.84 & -0.92 & 0.08 &    O8 V     & 230 & 186 &   IMACS f/4  &  101222 \\
51036 & 15.10 & 15.16 & -0.87 & 0.71 &    Be$_1$        & -- & -- &     IMACS f/4  &  101222 \\
51214 & 14.72 & 14.76 & -1.03 & 0.87 &    Be$_3$       & -- & -- &     IMACS f/4  &  081006 \\
51234 & 13.58 & 13.66 & -0.86 & 0.58 &    B1.5 III         & 187 & 187 &   IMACS f/4  &  070920 \\
51236 & 14.93 & 14.92 & -0.89 & 0.89 &    Be$_{4+}$       & -- & -- &     IMACS f/4  &  090824 \\
51373 & 13.72 & 13.79 & -1.08 & 0.84 &    O8 IIIze$_3$        & -- & -- &     IMACS f/4  &  090824 \\
51384 & 13.26 & 13.42 & -0.92 & 0.44 &    O9.5 Ia          & 82 & 123 &   IMACS f/4  &  070920 \\
51419 & 15.15 & 15.29 & -0.90 & 0.45 &    O9 III   & -- & -- &     IMACS f/4  &  060912 \\
51424 & 14.99 & 15.02 & -0.94 & 0.82 &    B1e$_{3+}$    & 151 & 153 &   IMACS f/4  &  081007 \\
51435 & 14.92 & 15.12 & -0.89 & 0.29 &    O9 III-V         & 195 & 189 &   IMACS f/4  &  090824 \\
51500 & 14.72 & 14.97 & -0.93 & 0.15 &    O6 V     & 171 & 171 &   IMACS f/4  &  101221 \\
52257 & 14.57 & 14.61 & -0.87 & 0.77 &    B0e$_3$    & -- & -- &     IMACS f/4  &  090824 \\
52363 & 14.80 & 14.98 & -0.97 & 0.38 &    O7 III((f))e$_1$      & 170 & 170 &   IMACS f/4  &  101222 \\
52410 & 13.69 & 13.75 & -1.07 & 0.85 &    O8 III: ze$_3$ pec   & -- & -- &     IMACS f/4  &  090824 \\
52865 & 14.48 & 14.56 & -1.12 & 0.85 &    B0-B0.5e$_4$       & 192 & 181 &   MIKE  &  111025 \\
52959 & 15.03 & 15.25 & -0.87 & 0.14 &    B0 V     & -- & -- &     MIKE  &  111026 \\
53042 & 14.67 & 14.85 & -0.86 & 0.28 &    O9.5 V   & 154 & 163 &   IMACS f/4  &  090824 \\
53319 & 14.66 & 14.90 & -0.87 & 0.15 &    B0 V     & 173 & 173 &   IMACS f/4  &  081007 \\
53360 & 15.04 & 15.09 & -0.95 & 0.77 &    O7-9p:e$_2$    & -- & -- &     IMACS f/4  &  101222 \\
53480 & 13.99 & 14.24 & -1.15 & 0.03 &    --       & -- & -- &     --  &      -- \\
54456 & 13.35 & 13.55 & -0.88 & 0.25 &    B0 III           & 138 & 161 &   IMACS f/4  &  101222 \\
54721 & 12.69 & 12.88 & -0.95 & 0.35 &    O9 III   & 118 & 151 &   IMACS f/4  &  081007 \\
55417 & 14.58 & 14.62 & -0.99 & 0.83 &    B1e$_{4+}$           & -- & -- &     IMACS f/4  &  070920 \\
55952 & 14.43 & 14.64 & -0.89 & 0.21 &    B0.2 V   & 251 & 196 &   IMACS f/4  &  081007 \\
56503 & 14.87 & 14.95 & -0.92 & 0.68 &    O9 Ve$_2$     & -- & -- &     IMACS f/4  &  070920 \\
56587 & 14.68 & 14.75 & -0.97 & 0.73 &    B0 V   & -- & -- &     IMACS f/2  &  110721 \\
56662 & 14.87 & 11.85 & -1.30 & 0.54 &    B       & 182 & 182 &   MIKE  &  111026 \\
57397 & 14.66 & 14.69 & -0.96 & 0.82 &    B1e$_{3+}$    & 158 & 163 &   IMACS f/4  &  070920 \\
58168 & 14.94 & 14.87 & -0.89 & 1.10 &    --       & -- & -- &     --  &      -- \\
58756 & 14.79 & 14.74 & -0.94 & 1.08 &    Be$_{3+}$        & -- & -- &     IMACS f/4  &  081011 \\
58864 & 15.14 & 15.14 & -0.91 & 0.90 &    B0e$_3$          & -- & -- &     IMACS f/4  &  070920 \\
58947 & 13.63 & 13.63 & -0.85 & 0.74 &    B1 III   & 156 & 164 &   IMACS f/4  &  081011 \\
59319 & 14.23 & 14.32 & -0.99 & 0.68 &    early Ope$_3$    & -- & -- &     IMACS f/4  &  081007 \\
59421 & 13.64 & 13.78 & -0.86 & 0.42 &    B1 III   & 155 & 146 &   IMACS f/4  &  081011 \\
59867 & 14.83 & 14.76 & -1.06 & 1.25 &    B0e$_{3+}$    & -- & -- &     IMACS f/4  &  070920 \\
59897 & 15.19 & 15.28 & -0.93 & 0.66 &    B0e$_3$     & 154 & 148 &   MIKE  &  111026 \\
59977 & 14.81 & 14.81 & -0.98 & 0.94 &    B0e$_1$          & -- & -- &     IMACS f/2  &  110721 \\
60439 & 13.73 & 13.94 & -0.90 & 0.22 &    O + B      & -- & -- &     MIKE  &  111026 \\
60460 & 14.74 & 14.98 & -0.88 & 0.09 &    O8.5 V   & 173 & 173 &   IMACS f/4  &  081011 \\
61039 & 14.87 & 15.11 & -0.85 & 0.09 &    O9.5 V   & 184 & 184 &   IMACS f/4  &  081007 \\
61543 & 15.11 & 15.19 & -0.85 & 0.61 &    B       & -- & -- &     MIKE  &  111025 \\
61842 & 13.38 & 13.47 & -0.87 & 0.59 &    B1 III   & -- & -- &     IMACS f/2  &  110719 \\
62416 & 14.59 & 14.68 & -0.92 & 0.61 &    O9 V     & 135 & 135 &   IMACS f/4  &  070920 \\
62638 & 14.57 & 14.67 & -0.98 & 0.65 &    O9.5 III-Ve$_2$      & -- & -- &     IMACS f/4  &  081007 \\
62661 & 13.01 & 12.98 & -0.85 & 0.91 &    B[e]     & 89 & 159 &   MIKE  &  111026 \\
62981 & 14.66 & 14.86 & -0.87 & 0.24 &    B0.2 V   & 160 & 166 &   IMACS f/4  &  070920 \\
63112 & 14.97 & 14.99 & -0.90 & 0.82 &     B0.5e$_{3+}$          & 113 & 159 &   IMACS f/4  &  081007 \\
63284 & 13.38 & 13.49 & -0.85 & 0.50 &    B1e$_{3+}$    & 124 & 161 &   IMACS f/4  &  070920 \\
63413 & 13.97 & 13.90 & -0.89 & 1.10 &    O9.5 Ia          & 144 & 163 &   IMACS f/4  &  081007 \\
63842 & 14.56 & 14.70 & -0.87 & 0.41 &    B1 V     & 141 & 141 &   IMACS f/4  &  081006 \\
63877 & 14.78 & 14.98 & -0.87 & 0.23 &    B0.2 V   & 129 & 166 &   IMACS f/4  &  070920 \\
64032 & 14.98 & 14.99 & -0.91 & 0.84 &    B1e$_{2+}$     & -- & -- &     IMACS f/4  &  090826 \\
64194 & 14.61 & 14.60 & -0.98 & 0.99 &    B0e$_3$     & 168 & 168 &   IMACS f/4  &  090826 \\
64453 & 13.41 & 13.38 & -1.06 & 1.14 &    B1e$_{3+}$       & 194 & 194 &   IMACS f/4  &  090826 \\
64710 & 14.51 & 14.56 & -0.85 & 0.68 &    B1.5e$_{3+}$          & 117 & 161 &   IMACS f/4  &  081007 \\
64773 & 15.05 & 15.25 & -0.87 & 0.23 &    O8.5 V   & 136 & 136 &   IMACS f/4  &  090826 \\
65103 & 15.03 & 15.09 & -0.97 & 0.74 &    Be$_{2+}$       & 125 & 125 &   IMACS f/4  &  081006 \\
65145 & 15.11 & 15.00 & -0.85 & 1.19 &    B1.5e$_2$          & 175 & 175 &   IMACS f/4  &  081006 \\
65318 & 14.78 & 14.95 & -0.89 & 0.35 &    O9 Ve$_1$          & 190 & 190 &   IMACS f/4  &  081007 \\
65346 & 14.99 & 15.19 & -0.99 & 0.36 &    O8 V     & 156 & 163 &   MIKE  &  111026 \\
65355 & 14.71 & 14.91 & -0.92 & 0.28 &    O8 V + O9        & -- & -- &     IMACS f/4  &  090826 \\
66160 & 12.92 & 13.09 & -0.92 & 0.40 &    O + B        & 125 & 166 &   IMACS f/4  &  081006 \\
66302 & 14.78 & 14.80 & -0.98 & 0.87 &    B1e$_{3+}$     & 174 & 174 &   IMACS f/4  &  081006 \\
66415 & 13.08 & 13.25 & -0.91 & 0.38 &    O9.7 Ia          & 166 & 166 &   IMACS f/4  &  081006 \\
66507 & 14.83 & 14.91 & -0.96 & 0.68 &    B1e$_{3+}$    & -- & -- &     IMACS f/4  &  081006 \\
67029 & 15.08 & 15.21 & -0.86 & 0.40 &    B1 V     & 148 & 141 &   IMACS f/4  &  081006 \\
67060 & 14.43 & 14.63 & -0.94 & 0.28 &    O7 Vz    & 126 & 164 &   IMACS f/4  &  090828 \\
67269 & 14.56 & 14.73 & -0.93 & 0.40 &    O7 V     & 144 & 140 &   IMACS f/4  &  090826 \\
67305 & 15.16 & 15.20 & -0.93 & 0.77 &    B1.5e$_3$           & 180 & 180 &   IMACS f/4  &  081006 \\
67334 & 14.93 & 15.05 & -0.94 & 0.53 &    B0 V     & 209 & 202 &   IMACS f/4  &  081006 \\
67673 & 14.80 & 14.84 & -1.00 & 0.83 &    O9.5 V:e$_2$         & -- & -- &     IMACS f/4  &  090826 \\
67893 & 13.92 & 13.97 & -1.05 & 0.87 &    B0e$_{3+}$         & -- & -- &     IMACS f/4  &  081006 \\
68071 & 14.99 & 15.19 & -0.91 & 0.28 &    O9.5 III         & 177 & 177 &   IMACS f/4  &  081006 \\
68157 & 15.14 & 15.37 & -0.85 & 0.10 &    B0.2 V   & 204 & 201 &   IMACS f/4  &  081006 \\
68427 & 13.52 & 13.51 & -1.06 & 1.09 &    B0.2e$_4$        & 157 & 157 &   IMACS f/4  &  090826 \\
68621 & 14.41 & 14.64 & -0.94 & 0.20 &    O9.5 III         & 156 & 174 &   IMACS f/4  &  090828 \\
68756 & 14.30 & 14.48 & -0.94 & 0.34 &    O7.5 III:   & 205 & 205 &   IMACS f/4  &  090828 \\
68963 & 15.20 & 15.39 & -0.87 & 0.25 &    B0 V     & 163 & 163 &   IMACS f/4  &  081006 \\
69155 & 14.50 & 14.69 & -0.85 & 0.28 &    B0.2 III         & 209 & 206 &   IMACS f/4  &  090828 \\
69460 & 14.82 & 14.95 & -0.99 & 0.53 &    O6.5 III((f))e$_2$   & 130 & 146 &   IMACS f/4  &  090826 \\
69555 & 14.35 & 14.59 & -0.96 & 0.17 &    O6.5 V   & 223 & 197 &     IMACS f/4  &  090826 \\
69598 & 15.03 & 15.23 & -0.95 & 0.22 &    O9 V     & 175 & 175 &   IMACS f/4  &  090828 \\
69630 & 13.79 & 14.01 & -0.85 & 0.16 &    B1 II    & -- & -- &     IMACS f/4  &  081006 \\
69769 & 15.13 & 15.30 & -0.93 & 0.37 &    B1.5 V   & 122 & 131 &   IMACS f/4  &  081006 \\
70149 & 14.41 & 14.61 & -0.90 & 0.25 &    O9 V     & 155 & 155 &   IMACS f/4  &  090826 \\
70663 & 14.60 & 14.63 & -0.89 & 0.80 &    B1.5e$_1$           & -- & -- &     IMACS f/2  &  110720 \\
71002 & 13.75 & 13.97 & -0.93 & 0.23 &    O9.5 III         & 154 & 163 &   IMACS f/4  &  090826 \\
71409 & 14.91 & 14.88 & -1.05 & 1.14 &    Be$_3$       & -- & -- &     IMACS f/4  &  090826 \\
71652 & 14.44 & 14.65 & -0.88 & 0.18 &    B0.5e$_2$           & -- & -- &     IMACS f/4  &  090825 \\
71815 & 15.20 & 15.43 & -0.96 & 0.21 &    O8 V     & 172 & 172 &   IMACS f/4  &  090826 \\
72204 & 14.34 & 14.57 & -0.92 & 0.20 &    O8 V + O8 V      & -- & -- &     IMACS f/4  &  090826 \\
72208 & 12.41 & 12.54 & -0.89 & 0.48 &    B1 II    & 131 & 133 &   IMACS f/4  &  090828 \\
72210 & 14.47 & 14.48 & -1.02 & 0.99 &    B0e$_3$      & -- & -- &     IMACS f/4  &  090825 \\
72535 & 13.42 & 13.45 & -1.11 & 1.00 &    O8-9: IIIpe:$_1$     & 162 & 162 &   IMACS f/4  &  090825 \\
72656 & 14.85 & 14.99 & -0.85 & 0.41 &    B0-1.5 III       & 135 & 168 &   IMACS f/4  &  090826 \\
72724 & 14.25 & 14.42 & -0.87 & 0.34 &    B0 V     & 201 & 201 &   IMACS f/4  &  090828 \\
72884 & 14.65 & 14.90 & -0.92 & 0.11 &    O9 V     & 222 & 204 &   IMACS f/4  &  090826 \\
72941 & 14.86 & 15.11 & -0.90 & 0.10 &    O9 V     & 110 & 126 &   IMACS f/4  &  090828 \\
73185 & 15.19 & 15.27 & -0.95 & 0.72 &    B        & -- & -- &     IMACS f/2  &  111012 \\
73337 & 12.96 & 13.12 & -0.96 & 0.47 &    O8.5 I((f))      & 202 & 202 &   IMACS f/4  &  090828 \\
73355 & 14.26 & 14.36 & -0.93 & 0.59 &    B0e$_2$    & -- & -- &     IMACS f/4  &  090825 \\
73795 & 14.68 & 14.73 & -1.03 & 0.82 &    mid Oe$_{3+}$       & -- & -- &     IMACS f/2  &  110720 \\
73913 & 13.57 & 13.63 & -0.91 & 0.72 &    O9.5 I   & 184 & 184 &   IMACS f/4  &  090825 \\
73952 & 12.51 & 12.62 & -0.88 & 0.52 &    B1.5 II          & 135 & 135 &   IMACS f/4  &  101221 \\
74828 & 12.97 & 13.08 & -0.86 & 0.52 &    Be       & -- & -- &     IMACS f/4  &  090825 \\
74946 & 15.13 & 15.14 & -0.99 & 0.88 &    B        & 73 & 131 &   IMACS f/4  &  101221 \\
75061 & 14.17 & 14.24 & -0.93 & 0.69 &    B1e$_2$    & 164 & 164 &   IMACS f/4  &  090825 \\
75126 & 14.65 & 14.39 & -1.14 & 2.02 &    O9 V     & 139 & 139 &   IMACS f/4  &  090825 \\
75210 & 14.57 & 14.67 & -0.89 & 0.59 &    O8.5 V           & 196 & 196 &   IMACS f/4  &  090825 \\
75626 & 14.87 & 15.14 & -0.93 & 0.04 &    O9 III-V         & 159 & 159 &   IMACS f/4  &  090825 \\
75689 & 14.31 & 14.34 & -1.08 & 0.94 &    Ope$_3$ pec    & -- & -- &     MIKE  &  111025 \\
75719 & 14.60 & 14.87 & -0.84 & 0.00 &    --       & -- & -- &     --  &      -- \\
75919 & 14.32 & 14.39 & -1.01 & 0.78 &    O9 IIIe$_1$         & 166 & 166 &   IMACS f/4  &  101221 \\
75980 & 14.12 & 14.17 & -1.02 & 0.81 &    B0e$_3$    & -- & -- &     IMACS f/4  &  090826 \\
75984 & 15.12 & 15.35 & -0.85 & 0.07 &    B0 III   & 229 & 212 &   MIKE  &  111026 \\
76253 & 15.04 & 15.27 & -0.90 & 0.13 &    B0.2 V + B    & 188 & 188 &   IMACS f/4  &  090825 \\
76371 & 13.76 & 13.96 & -0.91 & 0.30 &    O9.5 III         & -- & -- &     IMACS f/4  &  090826 \\
76654 & 14.74 & 14.75 & -1.17 & 1.04 &    Be$_3$       & -- & -- &     IMACS f/4  &  090826 \\
76657 & 15.14 & 15.31 & -0.94 & 0.43 &    O9-9.5 V         & 198 & 198 &   IMACS f/4  &  090826 \\
76773 & 13.46 & 13.59 & -0.86 & 0.45 &    Be       & -- & -- &     IMACS f/4  &  090825 \\
76870 & 14.92 & 14.98 & -0.99 & 0.77 &    Be       & -- & -- &     IMACS f/2  &  111013 \\
77290 & 15.10 & 15.10 & -1.04 & 1.00 &    B0.5e$_{2+}$          & -- & -- &     IMACS f/4  &  090825 \\
77368 & 13.76 & 14.00 & -0.93 & 0.15 &    O6 V     & 222 & 202 &   IMACS f/4  &  090826 \\
77458 & 13.00 & 13.15 & -0.92 & 0.45 &    B0.2e$_1$         & 204 & 204 &   IMACS f/4  &  090826 \\
77609 & 12.55 & 12.69 & -0.91 & 0.48 &    B0.5 I   & 192 & 192 &   IMACS f/4  &  090825 \\
77616 & 14.11 & 14.08 & -1.11 & 1.22 &    early Ope$_3$ pec            & -- & -- &     IMACS f/4  &  090826 \\
77734 & 14.87 & 14.93 & -0.97 & 0.74 &    --       & -- & -- &     IMACS f/4  &  090826 \\
77816 & 14.43 & 14.59 & -0.85 & 0.34 &    B0.2 III         & 168 & 168 &   IMACS f/4  &  090825 \\
77851 & 14.09 & 14.14 & -0.96 & 0.82 &    B0.2-1e$_{3+}$       & -- & -- &     IMACS f/4  &  090826 \\
78438 & 14.79 & 15.02 & -0.86 & 0.11 &    B0 V     & -- & -- &     IMACS f/2  &  110720 \\
78694 & 14.27 & 14.35 & -0.98 & 0.72 &    O8.5 IIIe$_{2+}$       & -- & -- &     IMACS f/2  &  110721 \\
79225 & 14.65 & 14.69 & -1.06 & 0.88 &    Be$_{4+}$       & 150 & 171 &   MIKE  &  111026 \\
79248 & 14.40 & 14.60 & -0.92 & 0.29 &    O8.5 V   & -- & -- &     IMACS f/2  &  110721 \\
79326 & 14.90 & 14.94 & -0.95 & 0.80 &    Be       & -- & -- &     IMACS f/2  &  110719 \\
79513 & 14.71 & 14.88 & -0.97 & 0.42 &    Be       & -- & -- &     IMACS f/2  &  110719 \\
79587 & 14.98 & 15.13 & -0.87 & 0.38 &    B        & -- & -- &     IMACS f/2  &  111012 \\
79697 & 14.86 & 14.93 & -1.04 & 0.79 &    Be$_4$       & 219 & 201 &   MIKE  &  111026 \\
79976 & 14.74 & 14.74 & -1.01 & 1.01 &    Be$_3+$       & -- & -- &     IMACS f/2  &  110720 \\
80075 & 14.90 & 15.07 & -0.87 & 0.35 &    B        & -- & -- &     IMACS f/2  &  111012 \\
80352 & 14.23 & 14.28 & -1.08 & 0.90 &    Be       & -- & -- &     IMACS f/2  &  111013 \\
80412 & 13.33 & 13.46 & -0.94 & 0.53 &    B0.7 II          & 184 & 150 &   IMACS f/4  &  090828 \\
80545 & 13.36 & 13.53 & -0.89 & 0.36 &    B0.5 II          & -- & -- &     IMACS f/4  &  090828 \\
80573 & 14.38 & 14.48 & -0.85 & 0.54 &    B1 V     & 179 & 179 &   IMACS f/4  &  090827 \\
80579 & 14.66 & 14.81 & -0.84 & 0.38 &    B0.7 V   & 168 & 163 &   IMACS f/4  &  090827 \\
80582 & 14.40 & 14.48 & -0.92 & 0.66 &    B        & -- & -- &     IMACS f/4  &  090828 \\
80998 & 14.24 & 14.46 & -0.96 & 0.25 &    O8 V     & -- & -- &     IMACS f/2  &  110721 \\
81019 & 14.24 & 14.43 & -0.93 & 0.33 &    O9.5 V   & 162 & 162 &   IMACS f/4  &  090827 \\
81071 & 14.67 & 14.70 & -0.99 & 0.88 &    Be$_1$       & -- & -- &     IMACS f/2  &  110721 \\
81169 & 15.02 & 15.23 & -0.86 & 0.19 &    B0.2 V   & 178 & 178 &   IMACS f/4  &  090827 \\
81258 & 14.96 & 15.18 & -0.88 & 0.14 &    B0-1.5 V         & 247 & 174 &   IMACS f/4  &  090827 \\
81465 & 14.85 & 14.88 & -1.02 & 0.89 &    Be$_3$        & -- & -- &     IMACS f/4  &  090826 \\
81634 & 14.42 & 14.50 & -0.86 & 0.57 &    B1.5 Ve$_3$           & 177 & 181 &   IMACS f/4  &  090826 \\
81646 & 14.15 & 14.36 & -0.92 & 0.26 &    O8 V     & 185 & 185 &   IMACS f/4  &  090826 \\
81647 & 14.62 & 14.85 & -0.89 & 0.17 &    B0.2 V   & 168 & 168 &   IMACS f/4  &  090827 \\
81696 & 15.07 & 15.17 & -0.91 & 0.55 &    B1 V     & 194 & 187 &   IMACS f/4  &  090826 \\
81941 & 13.82 & 14.03 & -0.89 & 0.24 &    O9.5 III         & 200 & 187 &   IMACS f/4  &  090827 \\
82322 & 14.12 & 14.37 & -0.88 & 0.09 &    O9.5 III         & 171 & 171 &   IMACS f/4  &  090827 \\
82328 & 14.87 & 14.96 & -1.01 & 0.71 &    B0e$_2$     & -- & -- &     IMACS f/4  &  090826 \\
82387 & 15.07 & 15.26 & -0.87 & 0.27 &    Be       & -- & -- &     IMACS f/2  &  111011 \\
82408 & 14.39 & 14.59 & -0.87 & 0.23 &    B1 III   & 196 & 176 &   IMACS f/4  &  090826 \\
82444 & 15.15 & 15.30 & -0.95 & 0.44 &    B0 V     & 199 & 169 &   IMACS f/4  &  090826 \\
82489 & 14.17 & 14.22 & -1.04 & 0.86 &    O9: IIIpe$_{4+}$      & -- & -- &     IMACS f/4  &  090827 \\
82572 & 14.45 & 14.40 & -0.85 & 0.96 &    Be       & -- & -- &     IMACS f/2  &  111011 \\
82711 & 14.32 & 14.36 & -1.00 & 0.83 &    B1 V     & -- & -- &     IMACS f/4  &  090826 \\
82783 & 14.88 & 15.12 & -0.89 & 0.11 &    B0.5 V   & 160 & 160 &   IMACS f/4  &  090827 \\
83017 & 14.80 & 15.03 & -0.87 & 0.15 &    O9.5 III         & 191 & 181 &   IMACS f/4  &  090826 \\
83073 & 15.10 & 15.27 & -0.87 & 0.34 &    B0.7 V   & 158 & 215 &     IMACS f/4  &  090826 \\
83074 & 15.12 & 15.35 & -0.85 & 0.10 &    Be       & -- & -- &     IMACS f/2  &  111013 \\
83171 & 14.28 & 14.39 & -0.94 & 0.58 &    B0e$_2$    & -- & -- &     IMACS f/4  &  090826 \\
83202 & 12.61 & 11.53 & -0.87 & 0.47 &    Be       & -- & -- &     IMACS f/2  &  110719 \\
83224 & 14.60 & 14.65 & -1.12 & 0.90 &    B1e$_3$     & -- & -- &     IMACS f/4  &  090826 \\
83232 & 14.11 & 14.28 & -0.84 & 0.29 &    B1.5 III         & -- & -- &     IMACS f/4  &  090826 \\
83480 & 13.74 & 13.73 & -0.90 & 0.88 &    B[e]       & 164 & 164 &   IMACS f/4  &  090826 \\
83510 & 15.16 & 15.45 & -0.91 & 0.00 &    O8 V     & 153 & 159 &   IMACS f/4  &  090826 \\
83651 & 14.88 & 15.08 & -0.86 & 0.24 &    B       & -- & -- &     IMACS f/2  &  111012 \\
83678 & 12.99 & 13.18 & -0.95 & 0.36 &    O8.5 III         & 166 & 174 &   IMACS f/4  &  090826 \\
83962 & 14.52 & 14.61 & -0.96 & 0.65 &    Be$_2$        & -- & -- &     IMACS f/2  &  110720 \\
84277 & 14.42 & 14.57 & -0.85 & 0.38 &    B1e$_1$     & -- & -- &     IMACS f/2  &  110720 \\
84544 & 14.88 & 15.02 & -0.91 & 0.46 &    Be$_2$        & -- & -- &     IMACS f/2  &  111012 \\
\enddata
\tablenotetext{a}{This table is published in its entirety in the
  electronic edition of the $Astrophysical$ $Journal$.  A portion is
  shown here for guidance regarding its form and content.}
\tablenotetext{b}{From Massey (2002).}
\tablenotetext{c}{From Zaritsky et al. (2002).}
\tablenotetext{d}{Measured from Stanimirovi{\'c} et al. (1999).}
\tablenotetext{e}{Observed multiple times for binary monitoring; see Table 3.}
\tablenotetext{f}{From Massey (2012).}
\label{catalog}

\end{deluxetable}